\documentclass[12pt, preprint]{aastex}

\usepackage{lscape}
\usepackage{subfigure}

\newcommand{\kms}{\mbox{\,km\,s$^{-1}$ }}
\newcommand{\kmsn}{\mbox{\,km\,s$^{-1}$}}

\newcommand{\degree}{$^{\circ}$ }
\newcommand{\degn}{$^{\circ}$}

\newcommand{\CO}{$^{13}$CO }
\newcommand{\COJ}{$^{13}$CO J =  $ 1 \rightarrow 0 $ }
\newcommand{\COT}{$^{12}$CO }

\title{Kinematic Distances to Molecular Clouds identified in the Galactic Ring Survey}

\author{Julia Roman-Duval\altaffilmark{1}, James M. Jackson\altaffilmark{1}, Mark Heyer\altaffilmark{2}, Alexis Johnson\altaffilmark{1}, Jill Rathborne\altaffilmark{3}, Ronak Shah\altaffilmark{4}, Robert Simon\altaffilmark{5}}
\altaffiltext{1}{Institute for Astrophysical Research at Boston University, Boston MA 02215; jduval@bu.edu; jackson@bu.edu}
\altaffiltext{2}{Department of Astronomy, University of Massachusetts, Amherst, MA 01003-9305}
\altaffiltext{3}{Harvard Smithsonian Center for Astrophysics, Cambridge, MA 02138}
\altaffiltext{4}{MIT Lincoln Laboratory, 244 Wood Street, Lexington, MA 02134}
\altaffiltext{5}{Physikalisches Institut, Universitat zu Koln, 50937 Cologne, Germany }

\begin{abstract}
Kinematic distances to 750 molecular clouds identified in the \COJ Boston University-Five College Radio Astronomy Observatory Galactic Ring Survey (BU-FCRAO GRS) are derived assuming the Clemens rotation curve of the Galaxy. The kinematic distance ambiguity is resolved by examining the presence of HI self-absorption toward the \CO emission peak of each cloud using the Very Large Array Galactic Plane Survey (VGPS). We also identify 21 cm continuum sources embedded in the GRS clouds in order to use absorption features in the HI 21 cm  continuum to distinguish between near and far kinematic distances. The Galactic distribution of GRS clouds is consistent with a four-arm model of the Milky Way. The locations of the Scutum-Crux and Perseus arms traced by GRS clouds match star count data from the Galactic Legacy Infrared Mid-Plane Survey Extraordinaire (GLIMPSE) star-count data. We conclude that molecular clouds must form in spiral arms and be short-lived (lifetimes $<$ 10$^7$ yr) in order to explain the absence of massive, \CO bright molecular clouds in the inter-arm space.
\end{abstract}

\clearpage 
\begin{document}	
\maketitle

\section{Introduction}
\begin{figure}
\includegraphics[height=10cm]{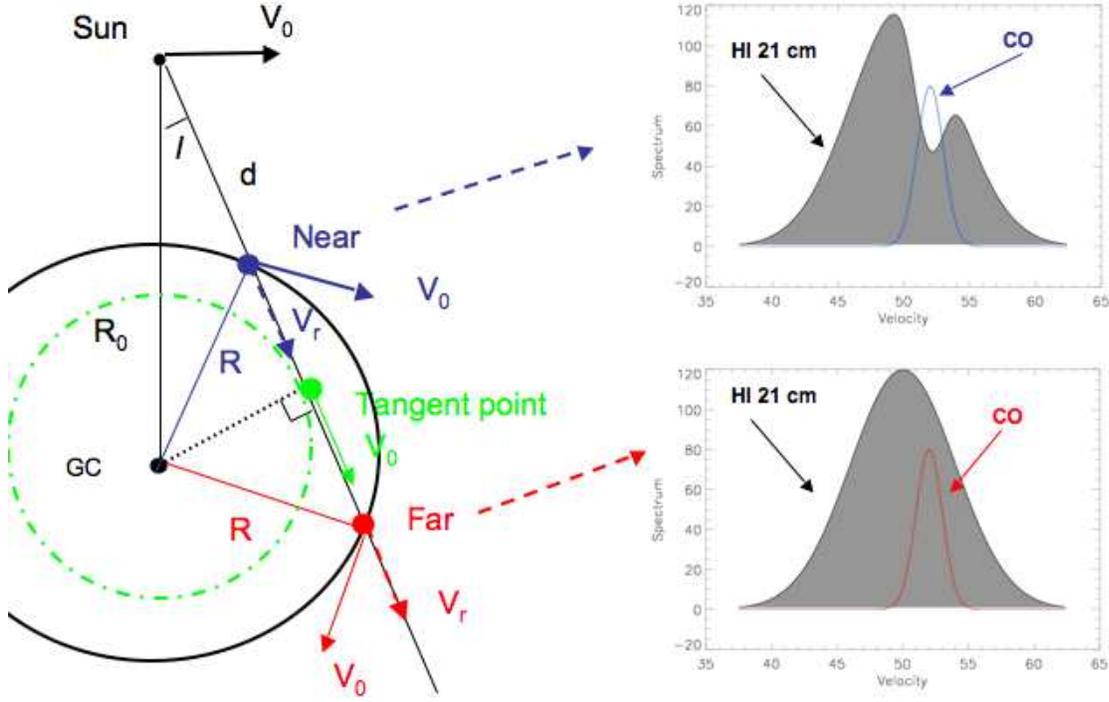}
\caption{Sketch of the HI self-absorption method to resolve the kinematic distance ambiguity. In the inner Galaxy, a single galactocentric radius (determined by the radial velocity of the cloud) corresponds to two distances along the line of sight, a near (in blue) and a far (in red) kinematic distance. The near and far kinematic distances correspond to the same radial velocity V$_r$, which is the projection of the orbital velocity V$_0$ of a cloud around the Galactic center onto the line of sight. At the tangent point, the orbital velocity of a cloud is parallel to the line of sight. In this case, the radial velocity is maximal and the near and far kinematic distances are identical. The cold HI embedded in a cloud located at the near kinematic distance absorbs the 21 cm radiation emitted by a warm HI background located at the far distance. Consequently, the HI  21 cm spectrum toward a near cloud exhibits an absorption line that is coincident with a \CO emission line from the cloud. A cloud located at the far distance does not lie in front of a warm HI background emitting at the same velocity as that of the cloud. Therefore, there is no absorption feature in the HI 21 cm spectrum toward a cloud located at the far kinematic distance.} 
\label{HISA_schem}
\end{figure}

Observations show that molecular clouds are the birthplace of stars. Theoretical models predict that molecular clouds form downstream of spiral shocks \citep[e.g.,][]{blitz80, dobbs06, dobbs08}. Turbulence and shocks cause diffuse, atomic gas to condense and cool mainly radiatively. Cold temperatures and an environment shielded from galactic radiation allow the formation of H$_2$ within 10 My \citep{bergin04, dobbs08}. Self-gravity subsequently triggers star formation through the gravitational instability when the gravitational energy of a fragment exceeds its internal (thermal, magnetic, turbulent) energy. From an observational point of view, O stars, OB associations, HII regions, and dust lanes are the best tracers of spiral arms in the Milky Way \citep[e.g.,][]{herbst75, lynds70} and other galaxies, because they trace recent star formation activity induced by a spiral perturbation. Previous studies of dense molecular clouds in our own Galaxy have also shown that stars form in molecular clouds \citep[e.g.,][]{beuther05, beuther07, pillai06,rathborne06, rathborne07, rathborne08a}. Thus, if our Galaxy is similar to other spiral galaxies, we should expect that molecular clouds in the Milky Way trace its spiral structure.  In order to deduce the spiral structure of the Milky Way from observations of molecular clouds, we need to derive (1) distances to these clouds and (2) a map of the distribution of molecular gas in our Galaxy. Distances to molecular clouds are also required to derive their fundamental properties such as mass, size, and density. However, the determination of distances to molecular clouds has proved to be a challenging task since they are not characterized by a typical length, nor by a typical luminosity. Consequently, the usual ``standard ruler'' or ``standard candle''  techniques do not apply.\\
\indent Several earlier investigations \citep[e.g.,][]{burton78,C85} used CO and HI spectral observations to derive a rotation curve for our Galaxy. They measured the maximum radial velocity of gas, which occurs for gas physically located at the tangent point, where the velocity vector of the cloud is aligned with the line of sight. Establishing the rotation curve was key to understanding the gas distribution in the Galaxy, since it was then possible to relate the spectroscopically observed radial velocity of a cloud to its galactocentric radius and distance. This method is called the kinematic distance method. Altough the kinematic distance method can be rendered quite inaccurate by localized velocity perturbations due to spiral shocks, expanding shells, and non-circular motions near the Galactic bar, it is the most efficient method to derive distances to molecular clouds.\\
\indent The Doppler-shift of a spectral line yields the radial velocity of the cloud, while the knowledge of the rotation curve of the Galaxy relates this velocity to a unique galactocentric radius. For a given longitude and radial velocity, there is a unique solution for the galactocentric radius given by the following expression:

\begin{equation}
r = R_0 sin(l) \: \frac{V(r)}{V_r + V_0 sin(l)}
\end{equation}

\noindent where R$_0$ is the galactocentric radius of the sun, V$_0$ is the orbital velocity of the sun around the Galactic center, V(r) is the rotation curve, and V$_r$ is the radial velocity of the cloud. In the inner Galaxy (r $<$ R$_0$), this galactocentric radius corresponds to two distances along the line of sight, the near and the far kinematic distances, located on either side of the tangent point (see Figure \ref{HISA_schem}). The two solutions for the distance are given by:

\begin{equation}
d = R_0 \: cos(l)  \pm \sqrt{r^2 - R_0^2 sin^2(l)}
\end{equation}

The kinematic distance ambiguity makes the determination of kinematic distances very challenging in the inner Galaxy. It stems from the fact that the radial velocity of a cloud, which is the projection of its orbital velocity around the Galactic center onto the line of sight, is the same at the near and far distances (see the top right panel of Figure \ref{cont_schem}). At the tangent point, the near and far distances are identical (d = R$_0$ cos (l) and r = R$_0$ sin(l)) and the orbital velocity of a cloud at this point is parallel to the line of sight. In this case, the radial velocity of a cloud is maximal and equal to its orbital velocity.  In the outer Galaxy (r $>$ R$_0$), there is a unique solution to the distance problem. The radial velocity decreases monotonically with distance to negative values.\\
\indent The kinematic distance ambiguity (KDA) has undoubtedly been a major obstacle to the understanding of the structure of the inner Milky Way. Earlier investigations of the spiral structure of the Milky Way using molecular data \citep[e.g.,][]{clemens88, gordon76, kolpak03, liszt84, sanders84, solomon79} have used various methods to resolve the kinematic distance ambiguity. For instance, \citet{clemens88}, who used the University of Massachusetts - Stony Brook \COT survey to probe the structure of the 5 kpc molecular ring, resolved the kinematic distance ambiguity by exploiting the different angular extent of the gas at the near and far kinematic distances. \citet{kolpak03} used 21 cm continnum absorption toward HII regions to resolve the kinematic distance ambiguity. This method requires an HII region to be embedded in the molecular cloud of interest. A consistent, systematic method to resolve the kinematic distance ambiguity for molecular clouds is thus needed. The method should be applicable to all molecular clouds, whether or not they contain a maser or an HII region.  \\
\indent In this paper, we present distances to molecular clouds identified in the Boston University - Five College Radio Astronomy Observatory (BU-FCRAO) Galactic Ring Survey (GRS) \citep{GRS, rathborne08}. We resolve the kinematic distance ambiguity by looking for HI self absorption (HISA) and 21 cm continuum absorption in the VLA Galactic Plane Survey (VGPS) \citep{vgps} toward the \CO emission peaks of the clouds. The HISA and 21 cm continuum absorption methods are described in section \ref{method}. In sections \ref{method}, 6, and \ref{methodology} we describe the data used for this analysis and the details of the methodology used to determine distances to the GRS clouds. Sections \ref{results}, \ref{map}, and \ref{discussion} present and discuss the results and their implications on the determination of the Milky Way's structure. The summary and conclusion are given in section \ref{conclusion}.

\section{HI self-absorption and 21 cm continuum absorption as a mean to the resolve the KDA}\label{method}
\begin{figure}
\includegraphics[height=12cm]{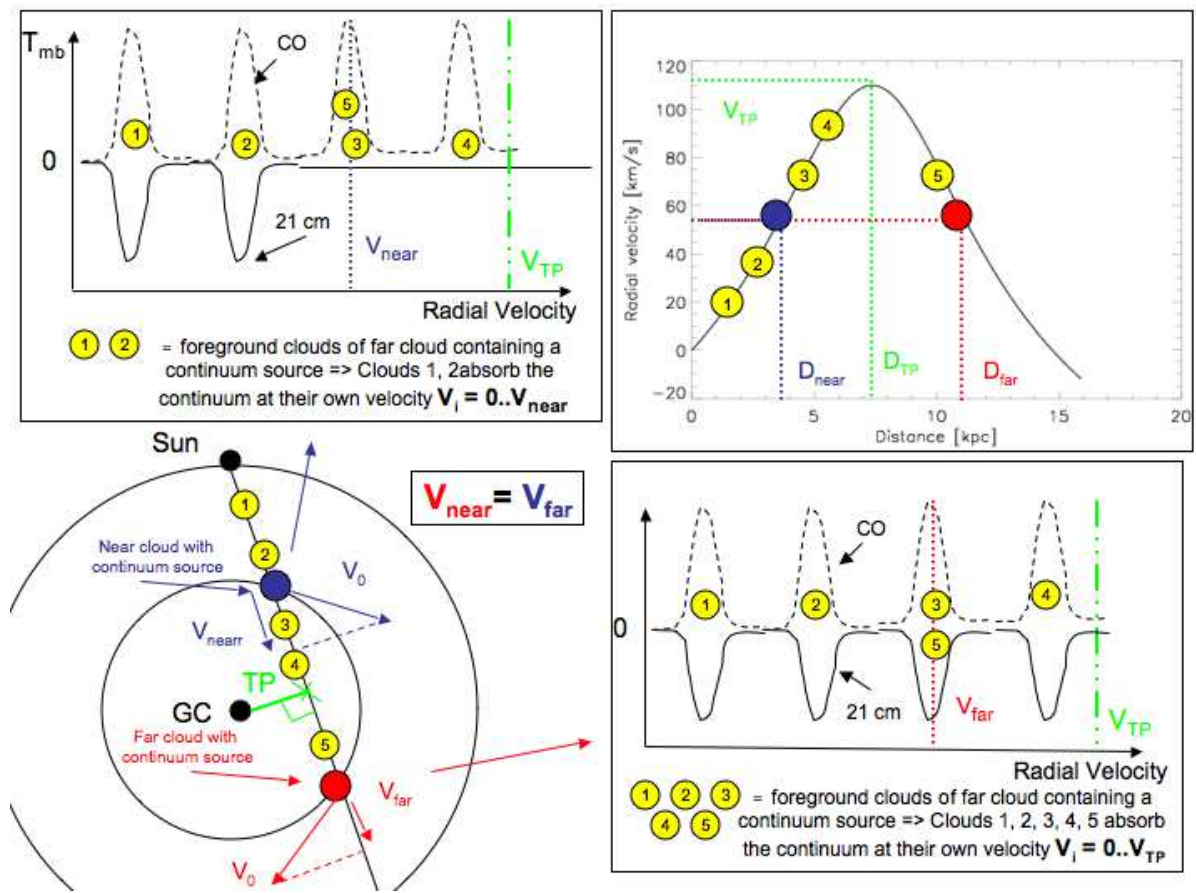}
\caption{Schematic of the 21 cm continuum absorption method to resolve the kinematic distance ambiguity.  The 21 cm continuum emitted by the source embedded in the cloud of interest is absorbed by all foreground molecular clouds. If the cloud of interest is located at the near kinematic distance (cloud in blue), the foreground molecular clouds (cloud 1 and 2) have velocities smaller than the velocity of the cloud (see top right panel).  As a consequence, the foreground molecular clouds absorb the 21 cm continuum emitted from the cloud of interest up to its radial velocity only. The HI 21 cm spectrum shows absorption lines at velocities up to the velocity of the cloud of interest (top left panel). On the other hand, if the cloud of interest is located at the far kinematic distance (cloud in red), the foreground molecular clouds (clouds 1, 2, 3, 4, 5) can have velocities up to the velocity of the tangent point. As a result, the HI 21 cm spectrum exhibits absorption lines up to the velocity of the tangent point, each absorption line corresponding to a \CO emission line from the foreground molecular clouds.  } 
\label{cont_schem}
\end{figure}

\indent The structure and composition of molecular clouds have been discussed by \citet{GL2006},  \citet{GL2007} and other authors. They showed that various molecules, such as $^{13}$CO and $^{12}$CO, coexist with molecular (H$_2$) and atomic (HI) hydrogen within the cold (T$_{kin}$ = 10 K) central regions of molecular clouds. The atomic hydrogen stems from both the remnant material before the cloud formed and predominantly from a continuous series of dissociative ionizations of H$_2$ by Galactic cosmic rays followed by recombinations. It can be shown that the HI density within dense molecular clouds does not depend on the overall gas density \citep{dyson80}.  Because this atomic hydrogen is shielded from outside radiation by the outer layers of the cloud, it is much colder than the rest of the Galactic atomic hydrogen, which has a temperature of 75-100 K and up to 10$^4$ K for the warm component \citep{kulkarni82}. Galactic cosmic rays constitute the only mechanism to heat the inner regions of molecular clouds, and explain their temperature of about 10 K. Several authors \citep[e.g.,][]{knapp74, FL2004,J04} showed that the high column densities and the low temperature  of HI in molecular clouds make molecular clouds sufficiently opaque and cold to allow the formation of absorption lines in the ubiquitous and warm background Galactic 21 cm radiation. \\
\indent  Absorption or self-absorption of the HI 21 cm line provides a technique to resolve the near/far kinematic distance ambiguity. The Inter-Stellar Medium (ISM) is filled with warm HI (T = 100 - 10$^4$ K), while cold (T = 10-30 K) HI is limited to dense molecular clouds. Furthermore, gas located at the near and far kinematic distances have the same radial velocity. Therefore, a molecular cloud located at the near kinematic distance lies in front of the warm HI background located at the far distance and emitting at the same velocity as that of the cloud. Because the HI embedded within the molecular cloud is much colder than the background warm HI, the radiation emitted at the far distance by the warm HI background is absorbed by the cloud on its way to Earth. Therefore, the HI 21 cm spectrum along the line of sight to a cloud located at the near distance will exhibit an absorption line that is coincident with a \CO emission line from the cloud. This phenomenon is known as HI self-absorption (HISA) and is explained in Figure \ref{HISA_schem}. A cloud located at the far kinematic distance lies in front a warm HI background, since warm HI fills the Galactic plane. However, this background has a radial velocity that is much different from the velocity of the cloud. This is due to the shape of the curve of the radial velocity versus distance shown in top right panel of Figure \ref{cont_schem}. All the warm HI at distances beyond the far cloud of interest are much lower than the velocity of the cloud. Therefore, the cloud cannot absorb the radiation emitted by the warm HI background. Furthermore, there is warm HI in the foreground of the cloud, emitting from the near distance at the same velocity as that of the cloud. Therefore, there is no HI self-absorption toward a cloud located at the far kinematic distance. Thus, an analysis of the HI 21 cm spectrum along the line of sight to a molecular cloud can resolve the kinematic distance ambiguity. If HI self-absorption is present at the same velocity as CO emission, then the molecular cloud is probably located at the near kinematic distance. If HISA is absent, the cloud is probably at the far kinematic distance.  \\
\indent In addition to the HISA method, absorption toward 21 cm continuum sources  (such as HII regions)  embedded within a molecular cloud can also be used as a distance probe. The brightness temperature of the 21 cm continuum radiation  from an HII region for instance is usually much higher than the temperature of the cold HI embedded in clouds. Therefore, the 21 cm continuum radiation emitted from a source embedded in the cloud of interest is absorbed by all the foreground molecular clouds along the same line of sight. If the cloud of interest, which contains the continuum source, is located at the near kinematic distance, the foreground molecular clouds all have velocities smaller than the velocity of the cloud. This can be seen in Figure \ref{cont_schem}, where the radial velocity increases monotonically with the near distance up to the tangent point. As a consequence, the foreground molecular clouds absorb the 21 cm continuum emitted from the cloud of interest up to its radial velocity only. Each absorption line in the 21 cm continuum should coincide with \CO emission from the foreground clouds. In this case, the HI 21 cm spectrum shows absorption lines at velocities up to the velocity of the cloud of interest, and each absorption line is associated with a \CO emission feature. On the other hand, if the cloud of interest is located at the far kinematic distance, Figure \ref{cont_schem} shows that the foreground molecular clouds can have velocities up to the velocity of the tangent point. In this case, the foreground molecular clouds absorb the 21 cm continuum radiation coming from the cloud of interest at velocities up to the velocity of the tangent point. As a result, the HI 21 cm spectrum exhibits absorption lines up to the velocity of the tangent point, each absorption line corresponding to a \CO emission line from the foreground molecular clouds. Whenever such 21 cm continuum sources are known to belong to the cloud under investigation, they can be used as a tool to determine distances. \\
   
\section{Data and identification of molecular clouds}\label{data}
\subsection{\CO data}
\indent Our sample contains molecular clouds that were identified by their \COJ  emission in the BU-FCRAO GRS \citep{GRS}. The GRS was conducted using the Five College Radio Astronomy Observatory (FCRAO) 14 m telescope in Salem, Massachussetts between 1998 and 2005. The survey, which used the SEQUOIA multipixel array, covers the range of Galactic longitude 18\degree $\leq$  $\ell$  $\leq$  55.7\degree and Galactic latitude $-$1\degree $\leq$  $b$  $\leq$  1\degn. The survey achieved a spatial resolution of 46" (sampling of 22") and a spectral resolution of 0.212 \kms for a noise variance $\sigma$ (T$_A^*$) = 0.13 K. The survey covers the range of velocity $-$5 to 135 \kms for Galactic longitudes $\ell$ $\leq$ 40\degree and $-$5 to 85 \kms for Galactic longitudes $\ell$ $\geq$  40\degn. \\
\indent The GRS is the first fully sampled large scale CO survey of the Galactic plane. Because of its resolution and sampling, and because the GRS uses \CO, which has an optical depth 50 times lower than that of \COT, it allows a clear detection and separation of molecular clouds both spatially and spectrally. Unlike for \COT surveys, the identification of molecular clouds in the GRS is not affected by velocity-crowding and line-blending.  \\
\indent Using the algorithm CLUMPFIND \citep{williams94} applied to the GRS smoothed to 0.1\degree spatially and to 0.6 \kms spectrally, 829 molecular clouds were identified by \citet{rathborne08}. CLUMPFIND identifies as clouds a set of contiguous voxels with values higher that a given threshold. We refer the reader to \citet{rathborne08} for the details of the identification procedure. The cloud parameters such as Galactic longitude, latitude and velocity of the \CO emission peak, velocity dispersion, angular extent, and antenna temperature were estimated by CLUMPFIND. \\

\subsection {H I data}

\indent The HI self-absorption analysis was performed using the VGPS \citep{vgps}. The survey mapped the HI 21 cm line in the range of Galactic longitudes 18\degree $\leq$  $\ell$  $\leq$  67\degree and latitude $-$1\degree $\leq$  $b$  $\leq$  1\degree. It achieved an angular resolution of 1', a spectral resolution of 1.56 \kms (FWHM) and a rms noise of 2 K per 0.824 \kms channel. Images of 21 cm continuum emission were made from channels without HI line emission. \\

\section{Kinematic distances to clouds: methodology}\label{methodology}
\begin{figure}
\includegraphics[height=10cm]{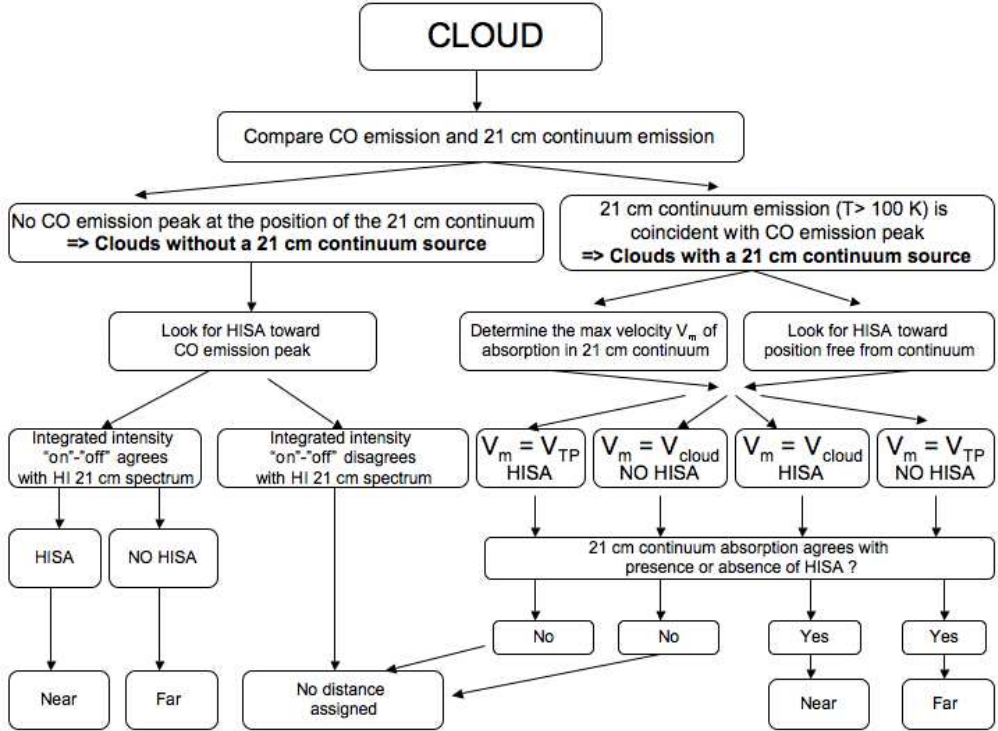}
\caption{Decision tree for the resolution of the kinematic distance ambiguity.  First, \CO  and 21 cm continuum emission were compared to determine whether the cloud contains a 21 cm continuum source. If a continuum source was present, the kinematic distance ambiguity was resolved using both absorption features in the continuum toward the source and HISA toward a position exhibiting strong \CO emission, but no 21 cm continuum emission. If both methods were in agreement, a distance was assigned. If not, no distance was assigned. For clouds that did not contain a 21 cm continuum source, only HISA toward the peak \CO emission was used to resolve the kinematic distance ambiguity. Only if the HI 21 cm spectrum and the ``on'' - ``off' HI 21 cm integrated intensity map agreed was a distance assigned.  } 
\label{decision_tree}
\end{figure}

\indent The near and far kinematic distances to each cloud in the GRS were calculated using the Clemens rotation curve of the Milky Way \citep{C85} scaled to (R$_0$, V$_0$)  = (8.5 kpc, 220 \kmsn), and included a small velocity correction (7 \kmsn) accounting for the measured solar peculiar motion. The kinematic distance ambiguity was resolved using the two methods described in section \ref{method}, depending on whether or not a 21 cm continuum source was embedded in the cloud of interest. The decision process is described in Figure \ref{decision_tree}. \\
\indent First, \CO  and 21 cm continuum emission were compared to determine whether the cloud contains a 21 cm continuum source. If no 21 cm continuum emission was detected over the projected area of the cloud, then HISA toward the \CO emission peak of the cloud of interest was used to resolve the kinematic distance ambiguity. In some cases, a 21 cm continuum source was detected in the most diffuse part of a cloud and was not coincident with a \CO emission peak. Because this case is likely a fortuitous alignement of a 21 cm continuum source and the cloud of interest, the 21 cm continuum source was not assigned to belong to the cloud in this case. The kinematic distance ambiguity was then resolved using HISA toward a position devoid of 21 cm continuum, but where strong \CO emission was present. \\
\indent If a 21 cm continuum source was detected within 0.02\degree of the \CO emission peak of the cloud of interest, then this source is most probably embedded in the cloud \citep{anderson09} and can be used as a distance probe. In this case, the kinematic distance ambiguity was resolved using both absorption features in the 21 cm continuum toward the source and HISA toward a position exhibiting strong \CO emission, but no 21 cm continuum emission. If both methods were in agreement, a distance was assigned. If not, no distance was assigned. The detailed methodology for each case is described in the next two paragraphs. 

\subsection {Resolution of the kinematic distance ambiguity for clouds that do not contain a 21 cm continuum source}\label{HISA_method}

\begin{figure}
\subfigure{
\includegraphics[height= 8cm]{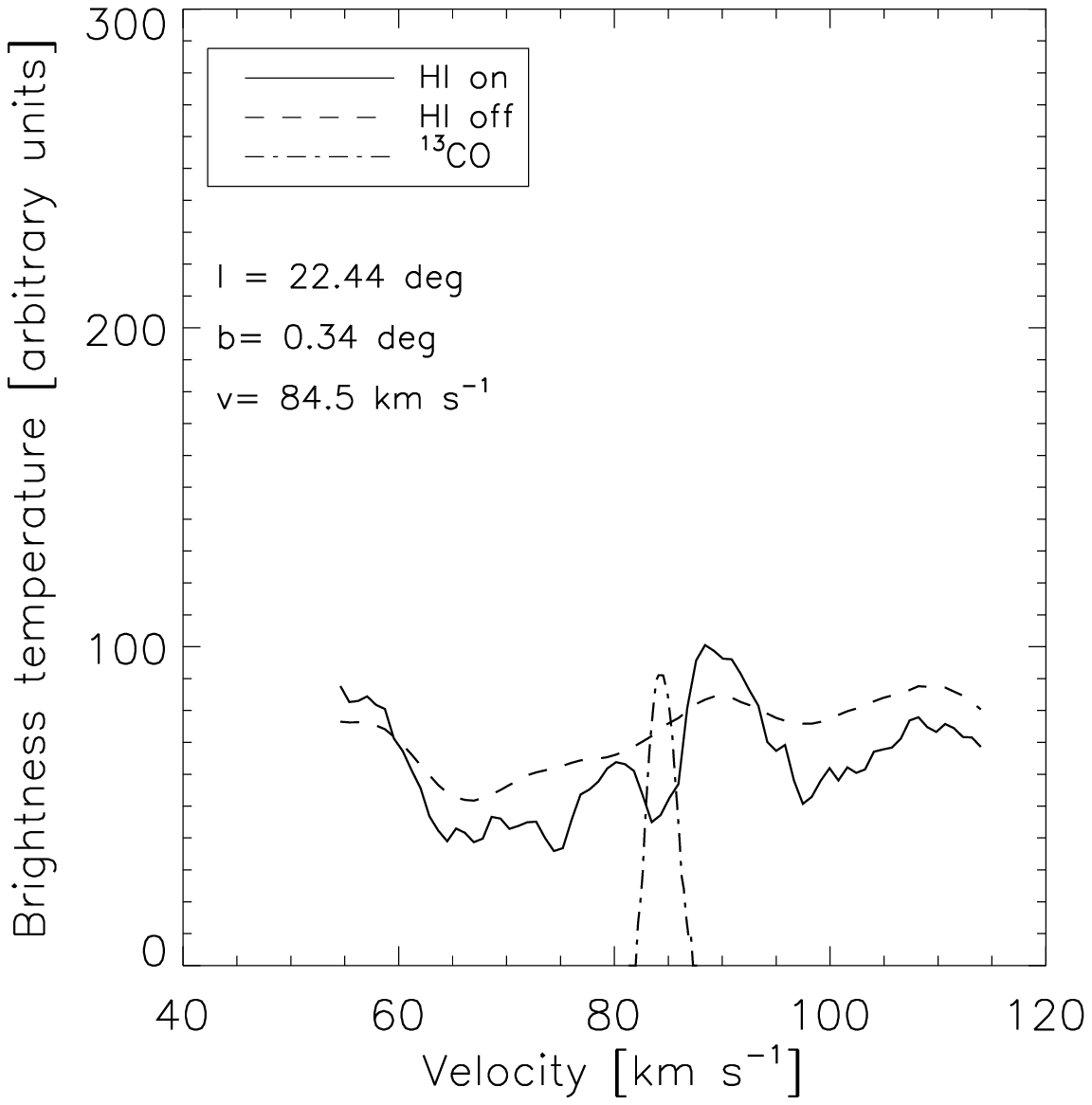}}
\hspace{0.1in}
\subfigure{
\includegraphics[height=8cm]{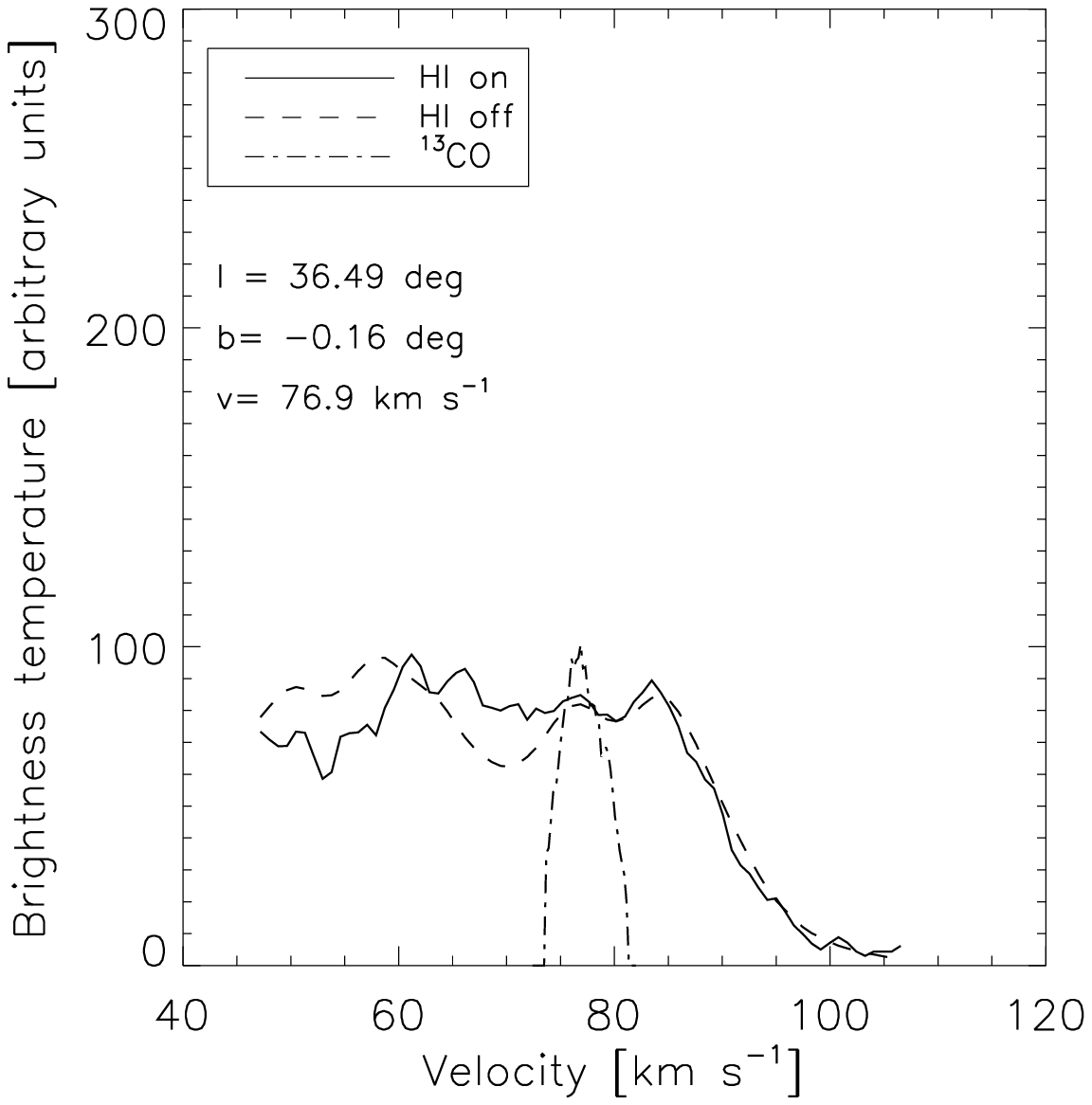}}
\caption{H I 21 cm spectra toward  ``on'' and ``off'' positions for near cloud GRSMC G021.41+00.31 (left) and far cloud GRSMC G036.50-00.13 (right). The ``on'' HI spectra were obtained toward the \CO emission peak. The HI 21 cm  ``off'' spectra were computed toward a position located less than 0.1 deg away from the \CO peak, but where  no \CO was detected (the integrated intensity at the ``off''  position was less than 3 $\sigma$ = 0.212 K \kmsn).  Overplotted are the \CO spectra for each cloud. We find that HISA is present toward GRSMC G021.41+00.31, implying that it is located at the near kinematic distance. In contrast, no HISA is found toward GRSMC G036.50-00.13, indicating that it is located at the far kinematic distance.}
\label{spec_near_far}
\end{figure}

\begin{figure}
\subfigure{
\includegraphics[height= 6cm]{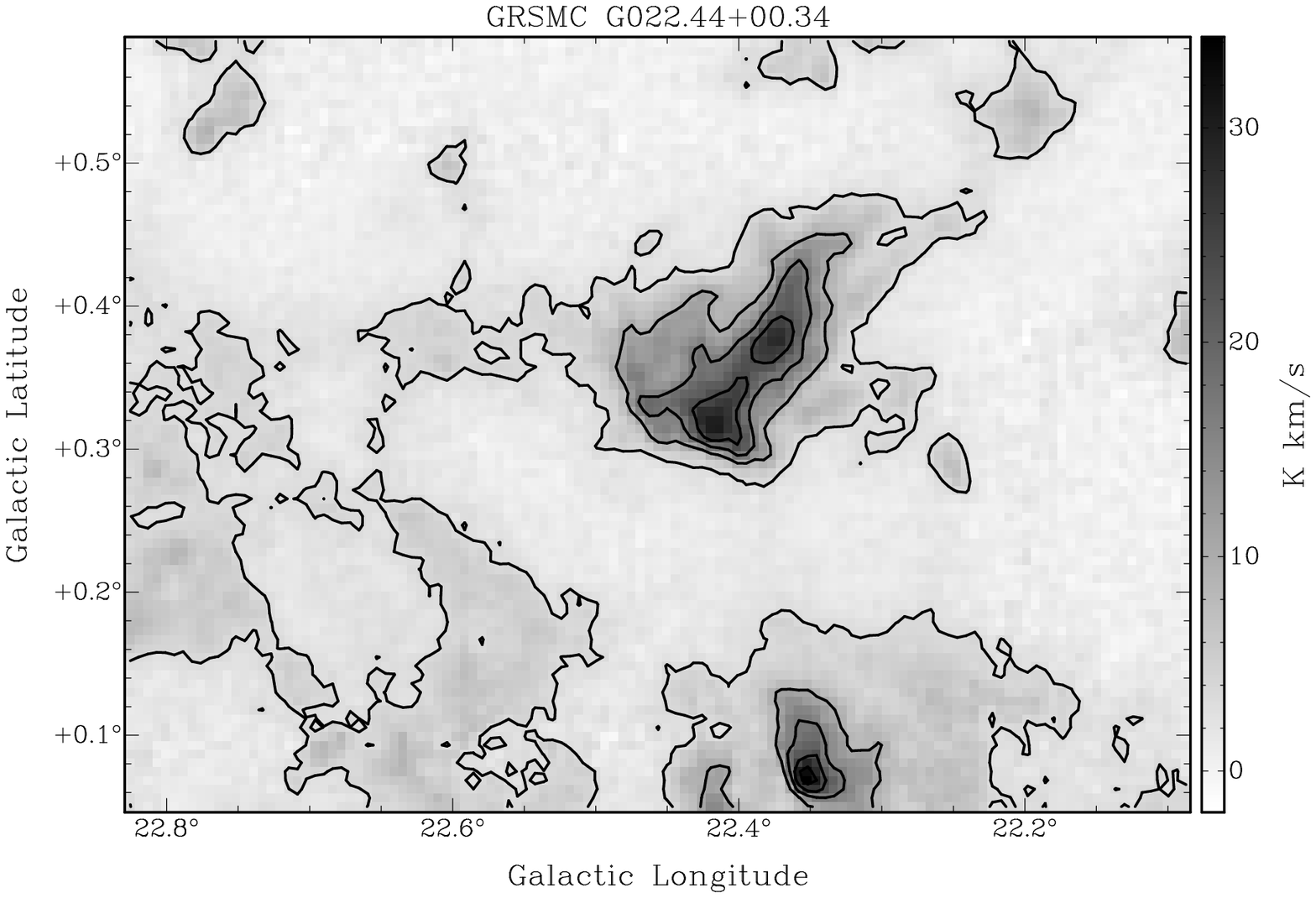}}
\vspace{0.1in}
\subfigure{
\includegraphics[height= 6cm]{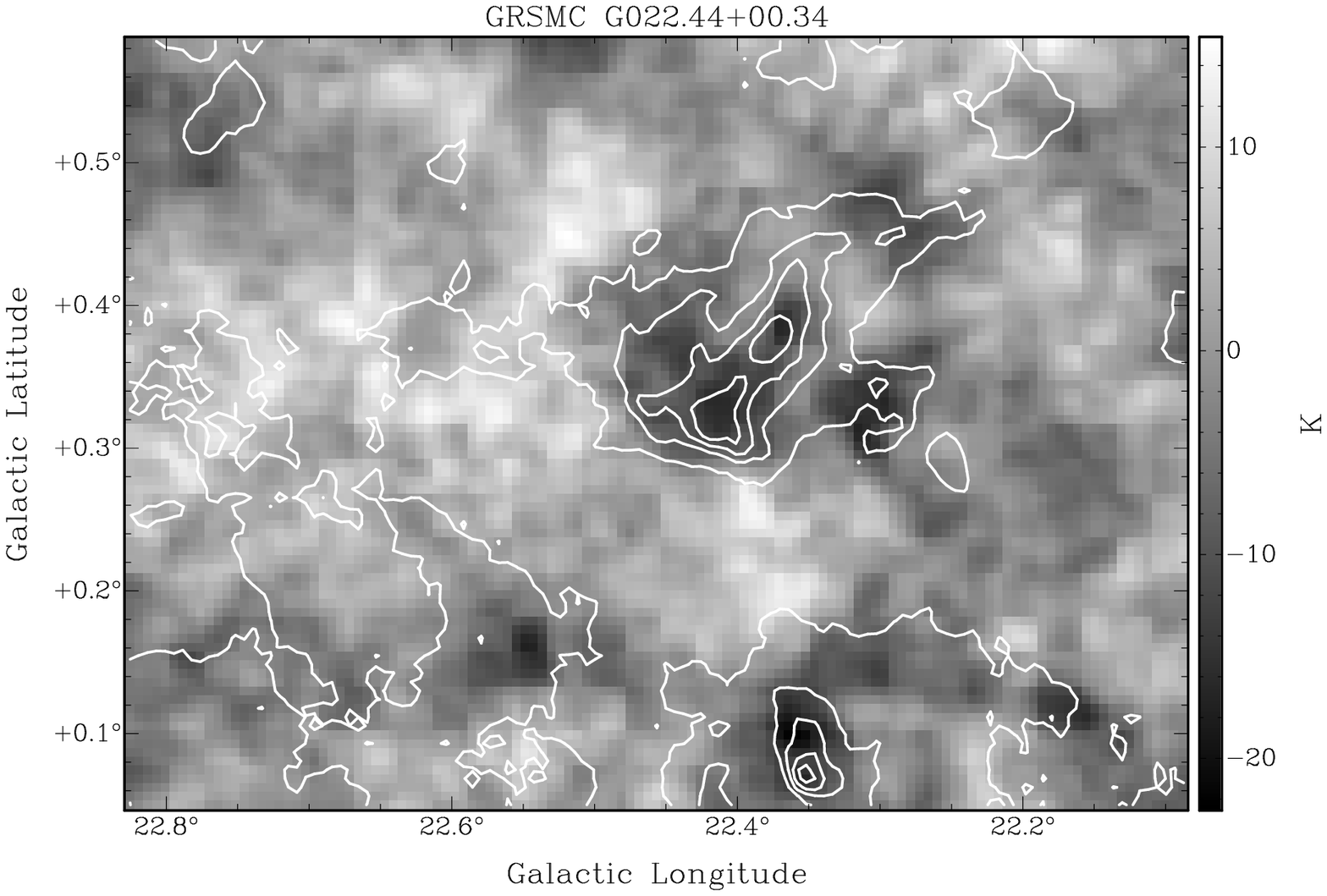}}
\caption{\CO integrated intensity (left) and HI 21 cm  ``on'' - ``off'' (right) maps for the cloud GRSMC G021.41+00.31. The HI ``on'' -``off'' integrated intensity image was obtained by subtracting the ``off'' map from the ``on'' map. The ``on'' map is integrated over the velocity range V $\pm$ $\Delta$V, where  V is the velocity of the cloud measured in the GRS, and $\Delta$V its FWHM velocity dispersion. The `off'' map was integrated over a 5 \kms range on each side of the line of the cloud. In the HI ``on'' - ``off'' map, the absorption feature represented by a dark patch matches well the morphology of the \CO emission, which is represented by contours on top of the ``on'' - ``off'' map as well as in the left panel. The contour levels are 10\%, 30\%, 50\%, 70\% and 90\% of the peak integrated intensity (34 K \kmsn). This suggests that the absorption line in the HI spectrum toward GRSMC G021.41+00.31 (see Figure \ref{spec_near_far}) is due to the presence of the cloud absorbing the background warm HI. Therefore, GRSMC G021.41+00.31 is located at the near kinematic distance.}
\label{intnear}
\end{figure}

\begin{figure}
\subfigure{
\includegraphics[height=6.5cm]{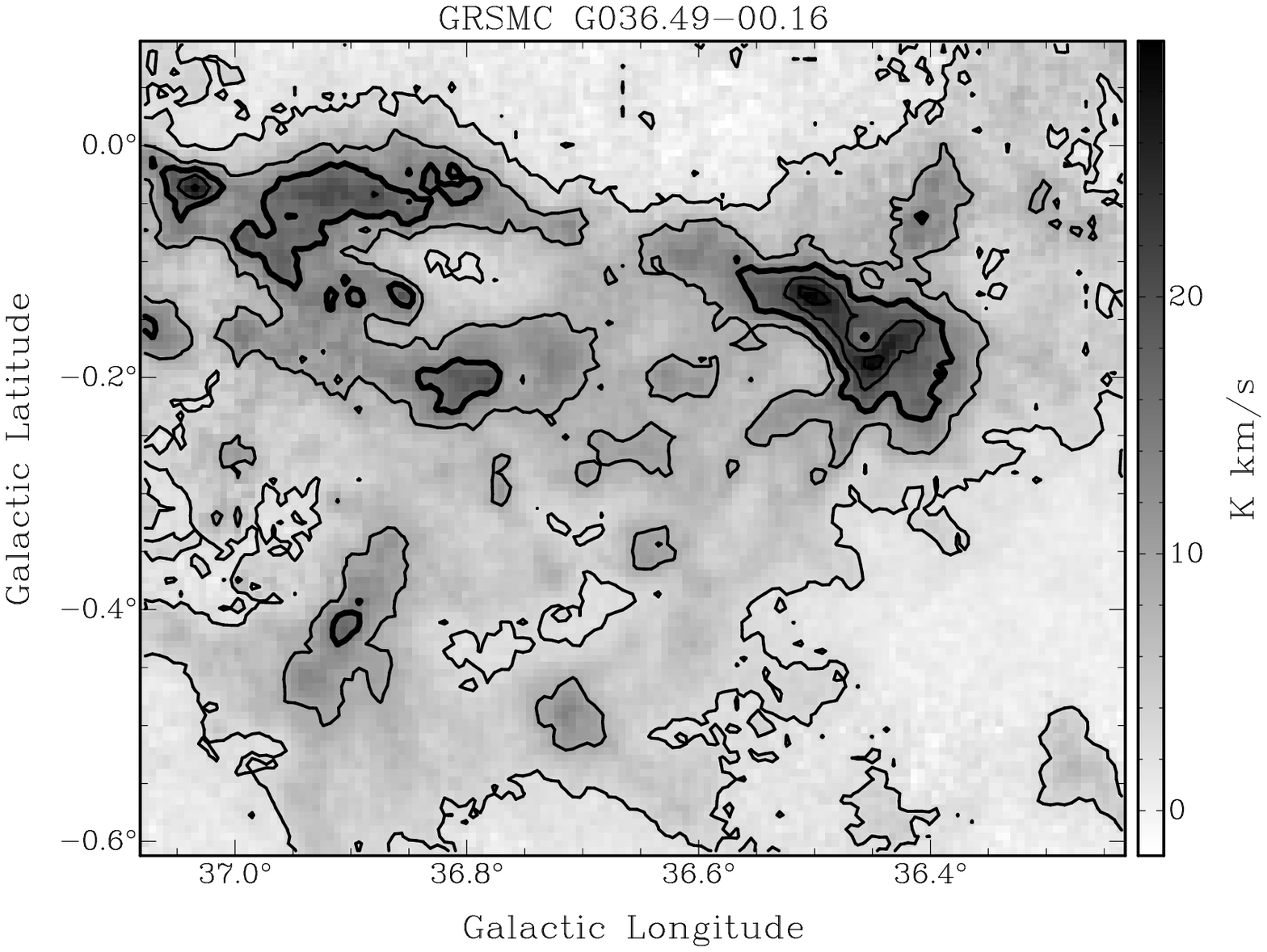}}
\subfigure{
\includegraphics[height=6.5cm]{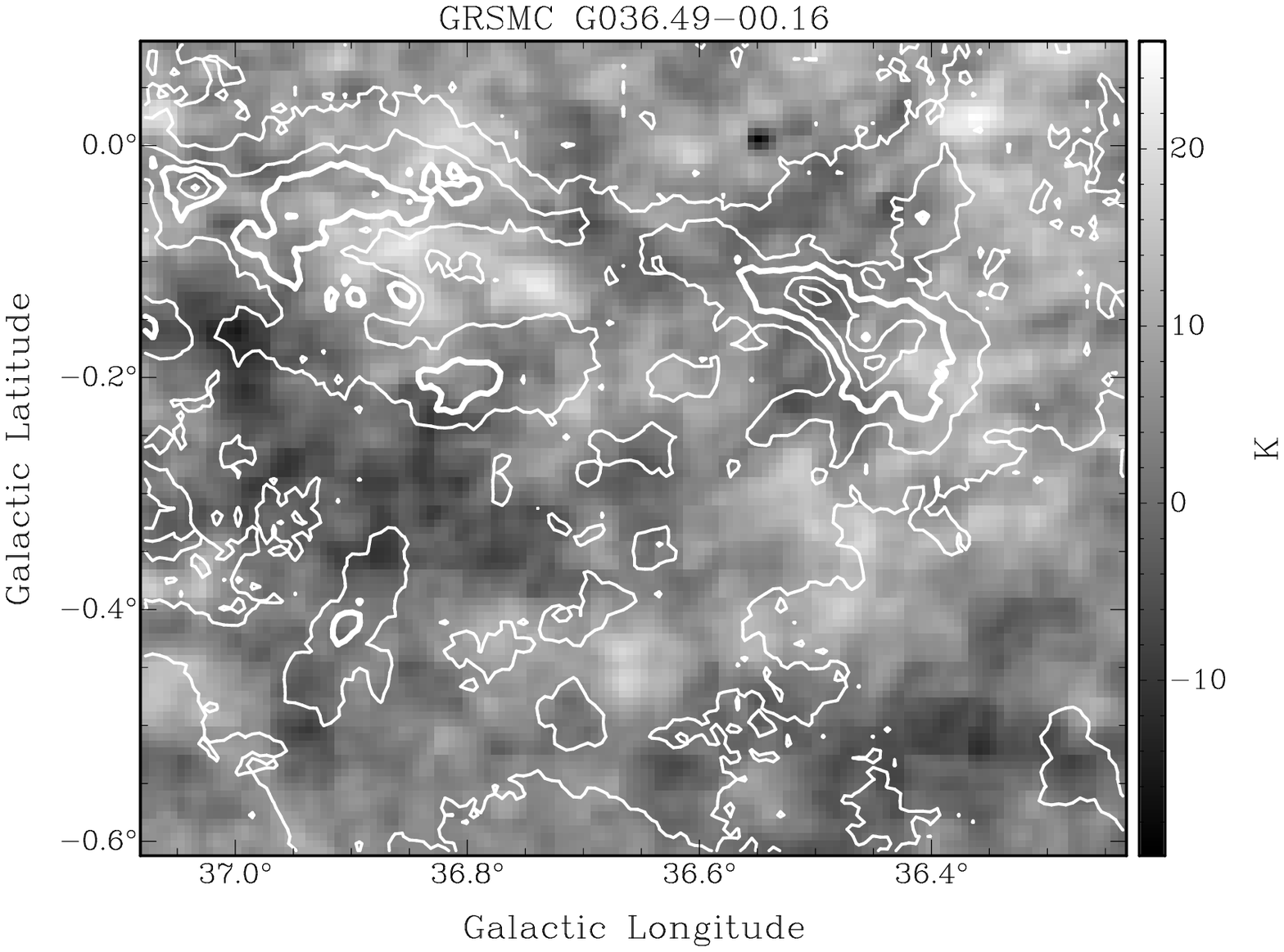}}
\caption{\CO integrated intensity (left) and HI 21 cm  ``on'' - ``off'' (right) maps for the cloud GRSMC G036.50-00.13. The HI ``on'' -``off'' integrated intensity image was obtained by subtracting the ``off'' map from the ``on'' map. The ``on'' map is integrated over the velocity range V $\pm$ $\Delta$V, where  V is the velocity of the cloud measured in the GRS, and $\Delta$V its FWHM velocity dispersion. The `off'' map was integrated over a 5 \kms range on each side of the line of the cloud. The \CO integrated intensity is shown as contours overplotted on the HI 21 cm ``on'' - ``off'' map and in the right panel. The contour levels are 10\%, 30\%, 50\%, 70\% and 90\% of the peak integrated intensity (30 K \kmsn). In the HI ``on'' - ``off'' map, there is no absorption feature (appearing as a dark patch in the ``on'' - ``off'' map) where the \CO emission is strong. This suggests that GRSMC G036.50-00.13 is located at the far kinematic distance. }
\label{intfar}
\end{figure}

\indent \CO and 21 cm continuum emission were first compared in order to assess whether the cloud of interest contains a 21 cm continuum source. The following applies to clouds toward which no 21 cm continuum emission was found (i.e., there was no 21 cm continuum source aligned with the \CO emission peak of the cloud). We examined HI self-absorption toward the \CO emission peak of each cloud using the HI 21 cm VGPS data. The HI 21 cm spectrum toward a molecular cloud located at the near distance exhibits an absorption line at the same velocity as \CO emission from the molecular cloud. There is no absorption feature in the HI 21 cm spectrum toward a molecular cloud located at the far kinematic distance (see Figure \ref{HISA_schem}). \\
\indent In order to detect the presence of HISA toward the GRS clouds, the \CO spectrum toward the peak \CO emission of each cloud from the GRS was first extracted. The presence of HISA was then examined by comparing the HI 21 cm spectrum along the same line of sight as the \CO emission peak to the HI 21 cm spectrum along  ``off'' positions. The ``off'' positions are located within 0.2\degree of the \CO emission peak of the cloud and are free from \CO emission at the velocity of the cloud. HISA is characterized by an absorption line in the ``on'' spectrum, which is coincident with the \CO line emission from the cloud, compared to the ``off'' spectrum. An example of this is displayed in Figure \ref{spec_near_far} for a near and a far cloud respectively. If HISA was present, the molecular cloud was assigned to the near kinematic distance. If it was absent, the cloud was assigned to the far kinematic distance. \\
\indent Without apriori knowledge of the exact shape of the HI spectrum, we cannot rule out that a dip in the HI 21 cm spectrum might be due to a fortuitous HI background fluctuation at the corresponding \CO velocity. To ascertain that HISA is present toward the cloud, we compared the morphology of the HI 21 cm absorption feature to the morphology of the \CO emission to ensure that the absorption feature detected in the HI 21 cm spectrum was correlated with \CO emission both spatially and spectrally. For this purpose, we derived maps of the HI integrated intensity ``on'' and  ``off'' the velocity of the cloud, and subtracted the ``off'' map from the ``on'' map.  The ``on'' maps were derived by integrating the HI 21 cm data cube over the velocity range V $\pm$ $\Delta$ V, where V is the radial velocity of the cloud measured in the GRS and $\Delta$V its FWHM velocity dispersion. The ``off'' maps were derived by integrating the HI 21 cm data cube over a 5 \kms range on both sides of the \CO emission line of the cloud, over the ranges V - $\Delta$V $\pm$ 2.5 \kms and V+ $\Delta$V $\pm$ 2.5 \kmsn. Those two ``off'' maps were averaged to form the final ``off'' maps. The \CO map was computed by integrating the \CO data over the velocity range of the cloud V $\pm$ $\Delta$V and by resampling the \CO data to the VGPS angular resolution and sampling. If a cloud is absorbing the background HI 21 cm radiation (i.e., the cloud is near), the absorption feature in the HI 21 cm spectrum, revealed in the HI 21 cm ``on'' - ``off'' maps by a  dark patch, should match morphologically the emission in the \CO map.  The comparison of the ``on'' - ``off'' maps to the \CO integrated intensity map thus confirmed or infirmed the presence of HISA. If HISA was detected or ruled out in both the HI 21 cm spectrum and the ``on'' - ``off'' integrated intensity maps, a distance was assigned accordingly. If the HI 21 cm spectrum and the ``on'' - ``off'' integrated intensity map were in contradiction, no distance was assigned. \\
\indent An example of a near (i.e., showing HISA) and a far (i.e., no HISA detected) cloud are given in Figures  \ref{intnear} and \ref{intfar} respectively. Based on this analysis, we resolved the kinematic distance ambiguity for 750 out of 829 clouds. The remaining 96 clouds are excluded from this analysis because either their distance determination was inconclusive or because they were found to be two clouds, one at the near distance blended together with a cloud along the same line of sight and at the same velocity, but at the far distance. The HISA analysis was decided inconclusive when the HI 21 cm spectrum and the ``on'' - ``off'' maps provided inconsistent results regarding the presence of HISA.

\subsection {Resolution of the kinematic distance ambiguity for clouds containing a 21 cm continuum source}

\begin{figure}
\centering
\includegraphics[height = 6cm]{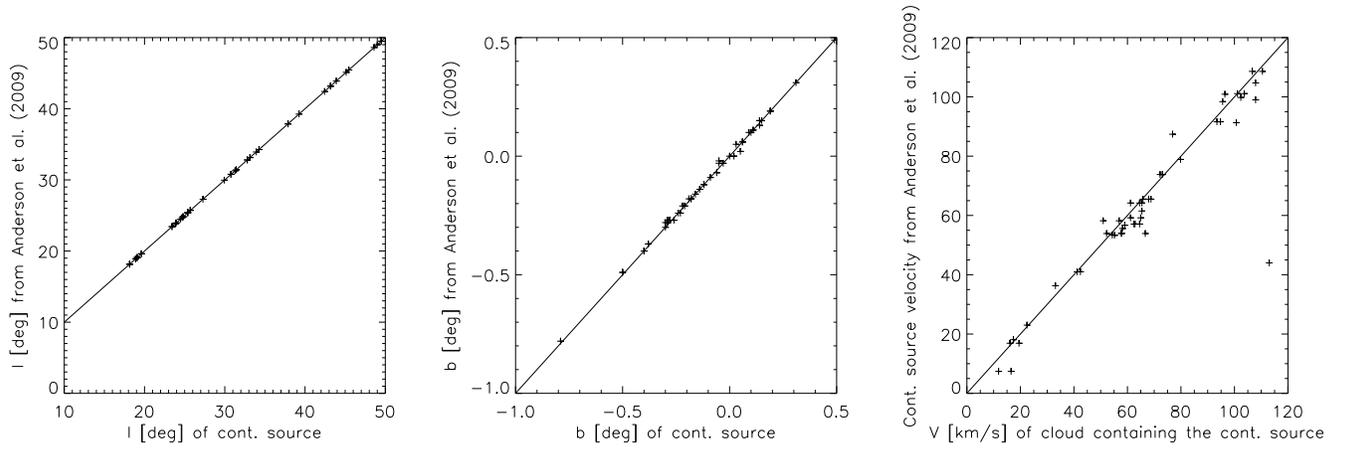}
\caption{Correspondance between the coordinates (Galactic longitude, latitude, and velocity) of the continuum sources identified in this paper and used as a distance probe (the velocity of the source being the same as the velocity of the parent cloud), and the coordinates of the closest match in the \citet{anderson09} catalog. Both sets of coordinates, and in particular the velocities, are in very good agreement, which validates the hypothesis that the 21 cm continuum sources that we identified are indeed embedded in their associated clouds.}
\label{id_hii_regions}
\end{figure}

\begin{figure}
\subfigure{
\includegraphics[height= 6cm]{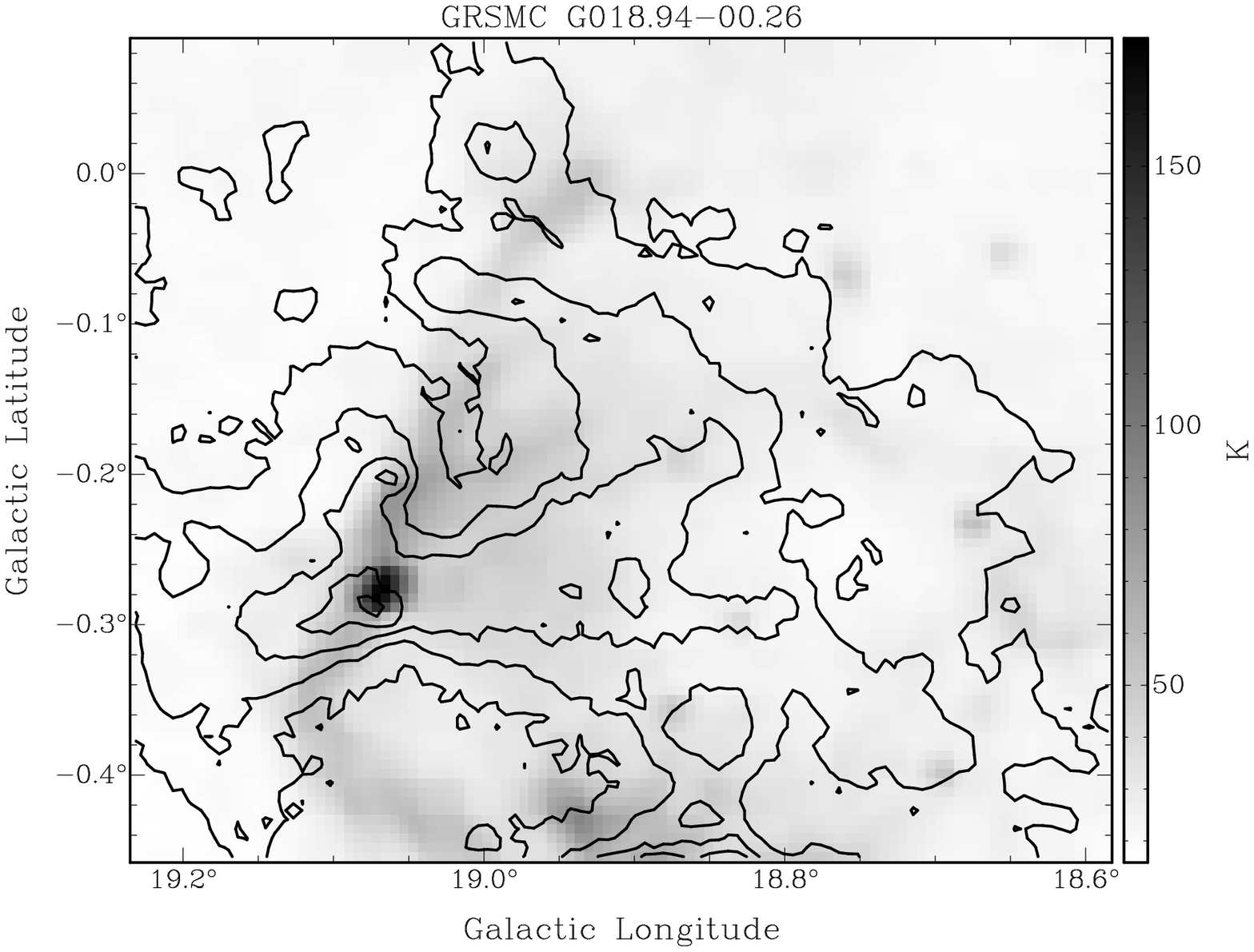}}
\subfigure{
\includegraphics[height= 6cm]{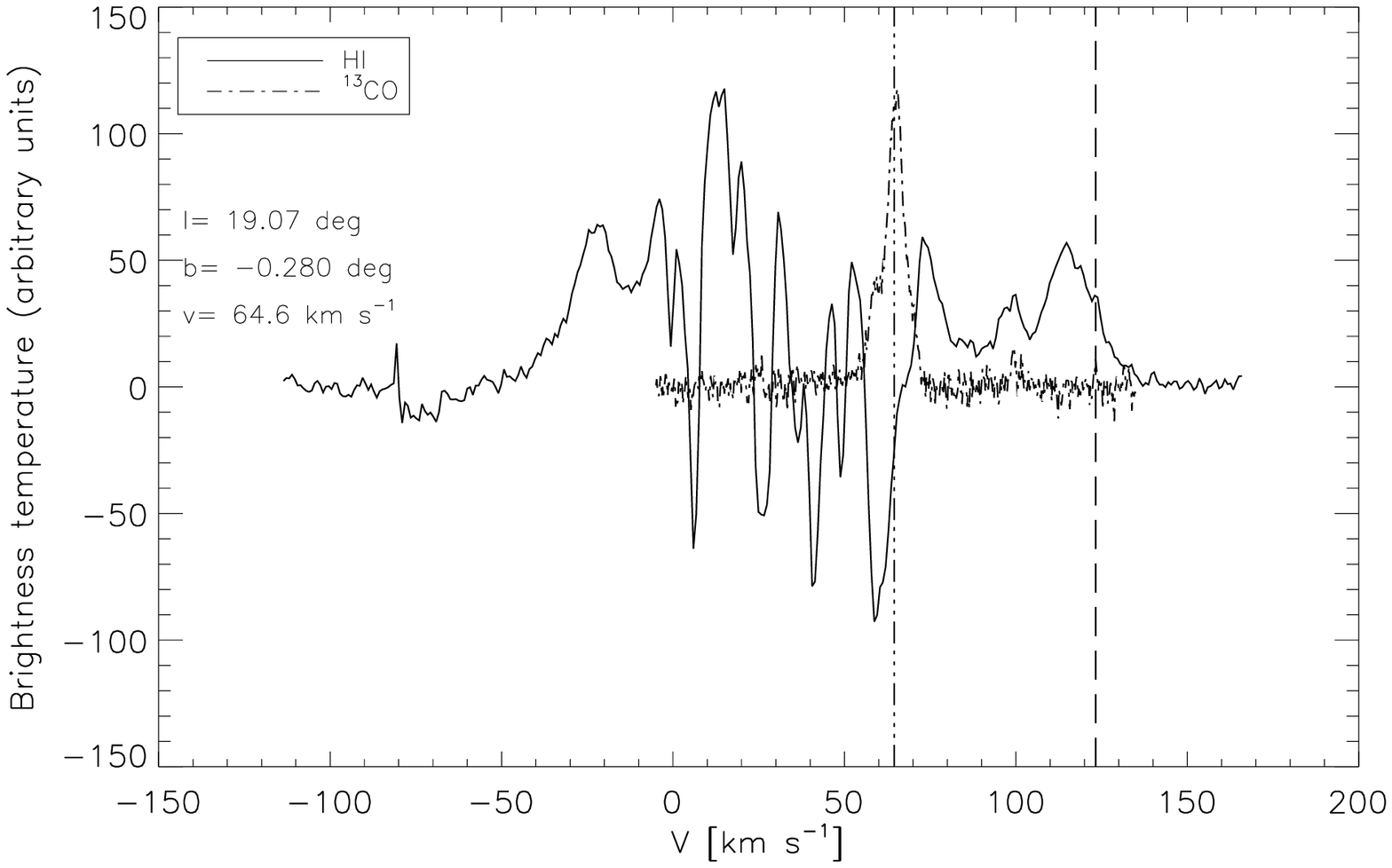}}
\caption{Left: \CO integrated intensity images as contours overplotted on the 21 cm continuum map for the cloud GRSMC G018.94-00.26. The contour levels are 10\%, 30\%, 50\%, 70\% and 90\% of the peak integrated intensity (35 K \kmsn). The colorbar indicates the brightness temperature of the continuum emission in K. The \CO and 21 cm continuum emission match morphologically. Therefore, continuum sources are embedded in GRSMC G018.94-00.26. Right: The right panel shows the HI 21 cm and \CO spectra toward the continuum sources embedded in GRSMC G018.94-00.26. The HI 21 cm spectrum was baseline-subtracted. The vertical dashed line indicates the tangent point, while the vertical dash-triple-dot line indicates the velocity of the cloud. The continuum toward GRSMC G018.94-00.26 is absorbed up to the velocity of the cloud (64.6 \kmsn) by foreground molecular cloud at velocities 2, 25, 40, and 50 \kmsn. There are no absorption lines in the 21 cm continuum at velocities greater than the velocity of GRSMC G018.94-00.26. Therefore, GRSMC G018.94-00.26 was assigned to the near kinematic distance.}
\label{cont_near_plots}
\end{figure}

\begin{figure}
\subfigure{
\includegraphics[height= 6cm]{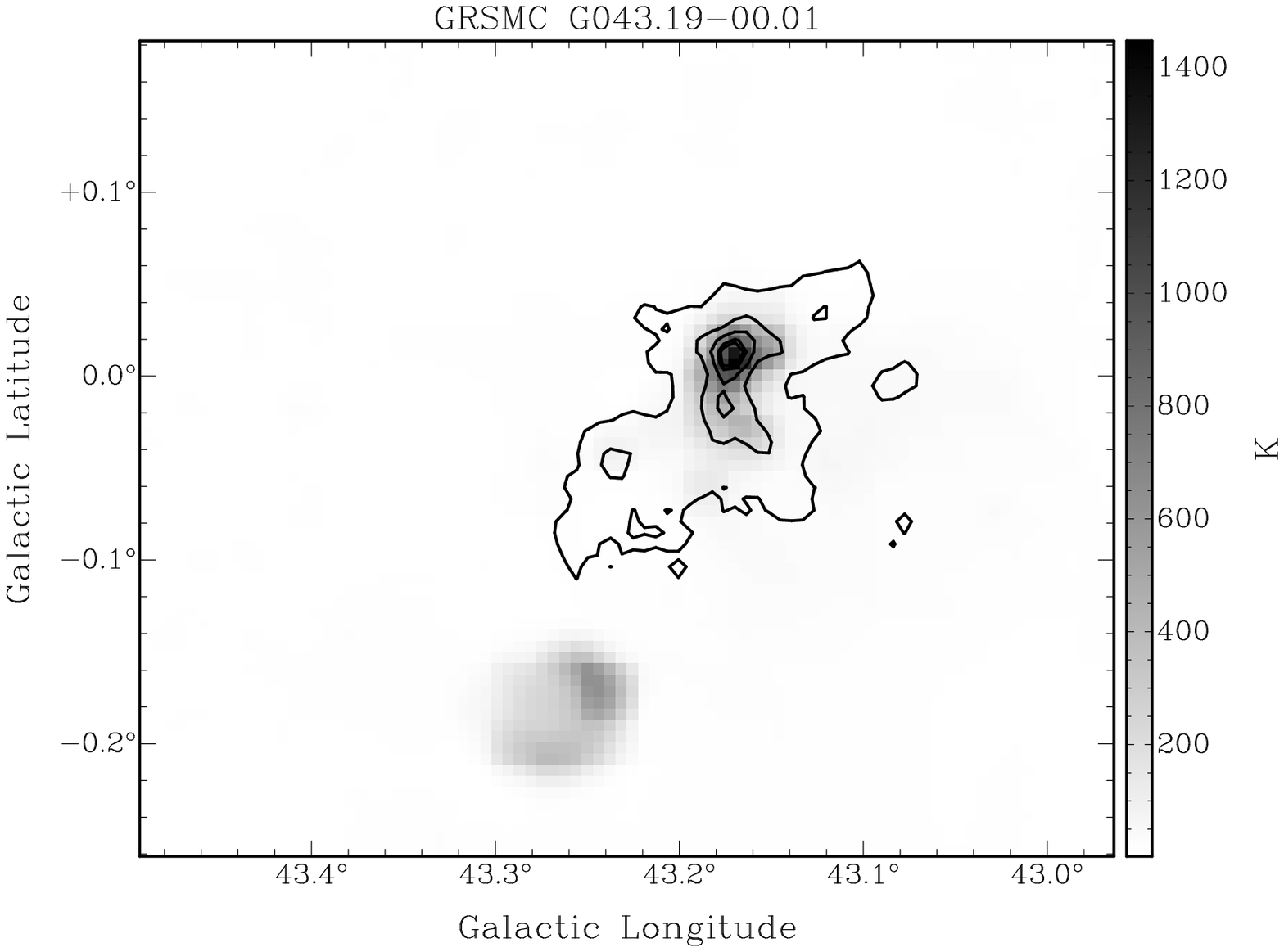}}
\subfigure{
\includegraphics[height=6cm]{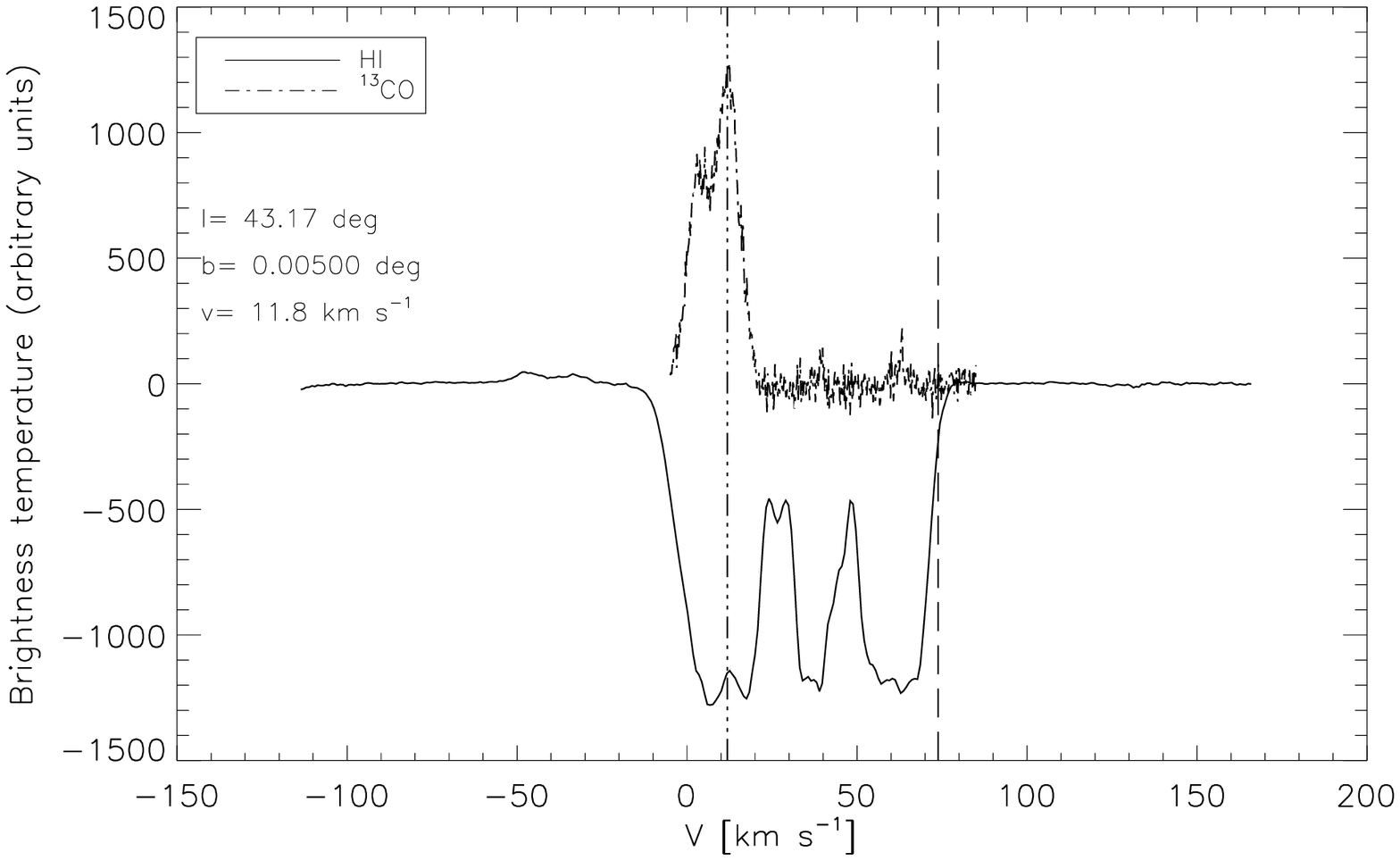}}
\caption{Left: \CO integrated intensity images as contours overplotted on the 21 cm continuum map for the cloud GRSMC G043.19-00.01 also called W49. The contour levels are 10\%, 30\%, 50\%, 70\% and 90\% of the peak integrated intensity (150 K \kmsn). The colorbar indicates the brightness temperature of the continuum emission in K. The \CO and 21 cm continuum emission match morphologically. Therefore, continuum sources are embedded in GRSMC G043.19-00.01. Right: The right panel shows the HI 21 cm and \CO spectra toward the continuum sources embedded in GRSMC G043.19-00.01. The HI 21 cm spectrum was baseline-subtracted. The vertical dashed line indicates the tangent point, while the vertical dash-triple-dot line indicates the velocity of the cloud. The 21 cm continuum is absorbed up to the velocity of the tangent point (75 \kmsn) by foreground molecular clouds with velocities greater than the velocity of GRSMC G043.19-00.01 (11.8 \kmsn). Therefore, GRSMC G043.19-00.01 was assigned to the far kinematic distance.}
\label{cont_far_plots}
\end{figure}

 \indent If one performs the HISA method on the line of sight to a 21 cm continuum source contained within a cloud, there is a risk that the HI gas associated with the cloud itself will absorb the continuum radiation, even if the cloud is at the far distance. Thus, there will be an absorption feature in the HI spectrum at the velocity of the cloud, which may be misinterpreted as HISA. Therefore, one has to be careful about using the HISA method when a continuum source is present. We checked that all the clouds analyzed with the method described in section \ref{HISA_method} were free from 21 cm continuum sources by looking at the VGPS continuum maps around the \CO emission peak associated with each cloud. Clouds containing a compact 21 cm continuum source with a peak brightness temperature T $>$ 100 K had their distances estimated with a separate analysis, which is described below. For these clouds, the HISA method described in section \ref{HISA_method} was also performed using the HI 21 cm spectrum along a line of sight with bright \CO emission, but no 21 cm continuum emission. We were then able to verify that the HISA and 21 continuum absorption methods gave the same result. Only if they did was a distance assigned.  \\
\indent The 21 cm continuum absorption method consists of examining the velocities of the different absorption features in the HI 21 cm spectrum along the line of sight to the continuum source embedded in the cloud of interest. Each absorption line is due to foreground molecular clouds absorbing the 21 cm continuum radiation from the source. In the first quadrant of the Galaxy, the radial velocity increases with distance up to the tangent point, where the velocity vector and the line of sight are parallel, and then decreases down to negative values (see top right panel of Figure \ref{cont_schem}). This is due to the shape of the rotation curve (mostly flat) and to the angle between the orbital velocity vector of the cloud around the Galactic center and the line of sight. If the cloud of interest lies at the near distance, only clouds with smaller radial velocities are located in front of it. Thus, the HI 21 cm spectrum (dominated by the continuum radiation) shows absorption lines coincident with \CO emission lines only at velocities smaller than the velocity of the cloud.  On the other hand, when the cloud containing the continuum source is at the far kinematic distance, the foreground clouds can have velocities up to the velocity of the tangent point. In this case, the HI spectrum will show absorption features at velocities greater than the velocity of the cloud of interest, up to the velocity of the tangent point. Each absorption line in the HI spectrum corresponds to emission lines at the same velocity in the \CO spectrum.  \\
\indent To apply this method as a distance probe, we need to ascertain that a continuum source detected in the 21 cm continuum VGPS maps is indeed embedded in the molecular cloud of interest. The correlation between strong \CO emission at the velocity of the cloud of interest and the position of the compact 21 cm continuum source confirms or infirms that the 21 cm continuum source is embedded in the cloud \citep{anderson09}. If the source is located within 0.02\degree of a \CO emission peak in the integrated intensity map of the cloud, it is most probably embedded in this cloud. We can also verify that the 21 cm continuum source is embedded in the cloud by using the H II region catalog from \citet{anderson09}, which contains radial velocities obtained from recombination lines. The velocities being obtained from recombination lines, they represent the velocities of the H II regions, independent of the parent cloud.  For each of the 21 cm continuum sources coincident with a CO emission peak, we assigned a galactic longitude and latitude measured in the continuum maps, and a velocity equal to the velocity of the "supposedly" parent cloud. We then searched the \citet{anderson09} catalog for the closest matches to each of our sources taking into account galactic longitude, latitude, and velocity. Both sets of coordinates were then compared. Figure \ref{id_hii_regions} shows the correspondance between the coordinates (Galactic longitude, latitude, and velocity) of the continuum sources identified in this paper and used as a distance probe (the velocity of the source being the same as the velocity of the parent cloud), and the coordinates of the closest match in the \citet{anderson09} catalog. If the continuum source is indeed embedded in the cloud, then the Galactic coordinates derived here should be identical to the Galactic coordinates from the \citet{anderson09} catalog. The velocity obtained from recombination lines (i.e., the velocity of the continuum source) should also be the same as the velocity of the cloud of interest. Overplotted on the figure is a 1:1 solid line. Both sets of coordinates, and in particular the velocities, are in very good agreement (except for one isolated, suspicious case). This validates the hypothesis that the 21 cm continuum sources that we identified are indeed embedded in their associated clouds since the velocity obtained from recombination lines is the same as the velocity of the cloud, and the galactic coordinates are identical.\\
\indent If a 21 cm continuum source was detected in the more diffuse part of the cloud, then it was not used to resolve the kinematic distance ambiguity. Instead, HISA was used toward a position devoid of 21 cm continuum emission, but exhibiting strong \CO emission.  \\
\indent The left panel of Figure \ref{cont_near_plots} shows a \CO integrated intensity image as contours, overplotted on a VGPS 21 cm continuum map for the cloud GRSMC G018.94-00.26. There is a good morphological match between the \CO and 21 cm continuum emission. Therefore, GRSMC G018.94-00.26 contains one or several non-resolved continuum sources, which we can use as a distance probe. The right panel of Figure \ref{cont_near_plots} shows the HI 21 cm and \CO spectra toward GRSMC G018.94-00.26. The cloud has a velocity of about 65 \kms as indicated by the vertical dash-triple-dot line, while the tangent point velocity, represented by the vertical dashed line, is about 125 \kmsn. The HI 21 cm spectrum shows absorption lines associated with \CO emission up to 65 \kmsn, but the \CO feature at 100 \kms is not associated with an absorption line in the HI 21 cm spectrum. Therefore, GRSMC G018.94-00.26 was assigned to the near kinematic distance (4.7 kpc). \\
\indent The left panel of Figure \ref{cont_far_plots} shows the morphological correlation between \CO emission and 21 cm continuum emission for the cloud GRSMC G043.19-00.00 (also called W49), which also contains continuum sources. The right panel of Figure \ref{cont_far_plots} shows the  HI  21 cm and \CO spectra toward GRSMC G043.19-00.00, which has a velocity of 11.8 \kms (vertical dash-triple-dot line). The tangent point velocity  for this longitude is 70 \kms (vertical dashed line). \CO emission lines are associated with absorption lines in the HI 21 cm spectrum for velocities up to the velocity of the tangent point. Therefore, GRSMC G043.19-00.00 (W49) was assigned to the far kinematic distance (11.4 kpc). The distance to W49 derived in this paper is consistent with the distance derived from proper motion of H$_2$O masers \citep{gwinn92}.

\section{Results}\label{results}
\subsection{Catalog}
\indent Distances to 750 molecular clouds were derived using the methods described in section \ref{method}. The final catalog of cloud distances is available online. As an example, Table \ref{grs_table} includes the position and distance of 30 GRS clouds. The first column contains the name of the cloud from \citet{rathborne08}. The second, third, and fourth columns contain the Galactic longitude (in degrees), Galactic latitude, and radial velocity (in \kmsn) respectively.  Columns 5, 6, and 7 contain the spatial FHWM of the cloud along Galactic longitude and latitude (in degrees) and its FWHM velocity dispersion respectively. The eighth column provides the kinematic distance to the cloud in kpc. Column 9 represents the \CO luminosities of the clouds in units of 10$^4$ K \kmsn pc$^2$. In order to compute the \CO luminosity of a cloud, an integrated intensity map of the cloud was first generated. The integrated intensity I($\ell$, $b$) was then summed over positions that exhibited values greater than 3$\sigma$ = 0.212 K \kmsn.  

\begin{equation}
L({^{13}CO}) = \left(\frac{d}{pc}\right)^ 2 \: \int_{I(\ell, b) \geq 0.2 K km \: s^{-1}} I(\ell, b) d\ell \: db
\end{equation}

\noindent Finally, the last column indicates whether a 21 cm continuum source was embedded in the cloud. If ``y'' is specified, then both absorption in the 21 cm continuum and HISA toward a position devoid of 21 cm continuum were used to resolve the kinematic distance ambiguity. If the column is left blank, then only HISA was used.\\
\indent  Table \ref{dist_stat} summarizes the statistics of the distance determination, with the number of near and far molecular clouds for which the kinematic distance ambiguity was resolved using HISA only or both HISA and absorption in the 21 cm continuum. The number of molecular clouds for which the kinematic distance ambiguity could not be resolved is also indicated for each cause of the indetermination: either the ``on'' - ``off'' HI 21 cm integrated intensity map was in disagreement with the HI 21 cm spectrum (HISA inconclusive), or the cloud was actually composed of two components at the same velocity, one being at the near distance and the other at the far distance, or the resolution of the kinematic distance ambiguity using absorption in the 21 cm continuum and HISA were in disagreement. Out of 829 molecular clouds identified in the GRS, we resolved the kinematic distance ambiguity for 750 molecular clouds.

\subsection{Sources of errors in the distance estimation}\label{errors}
\begin{figure}
\subfigure{\includegraphics[height=7cm]{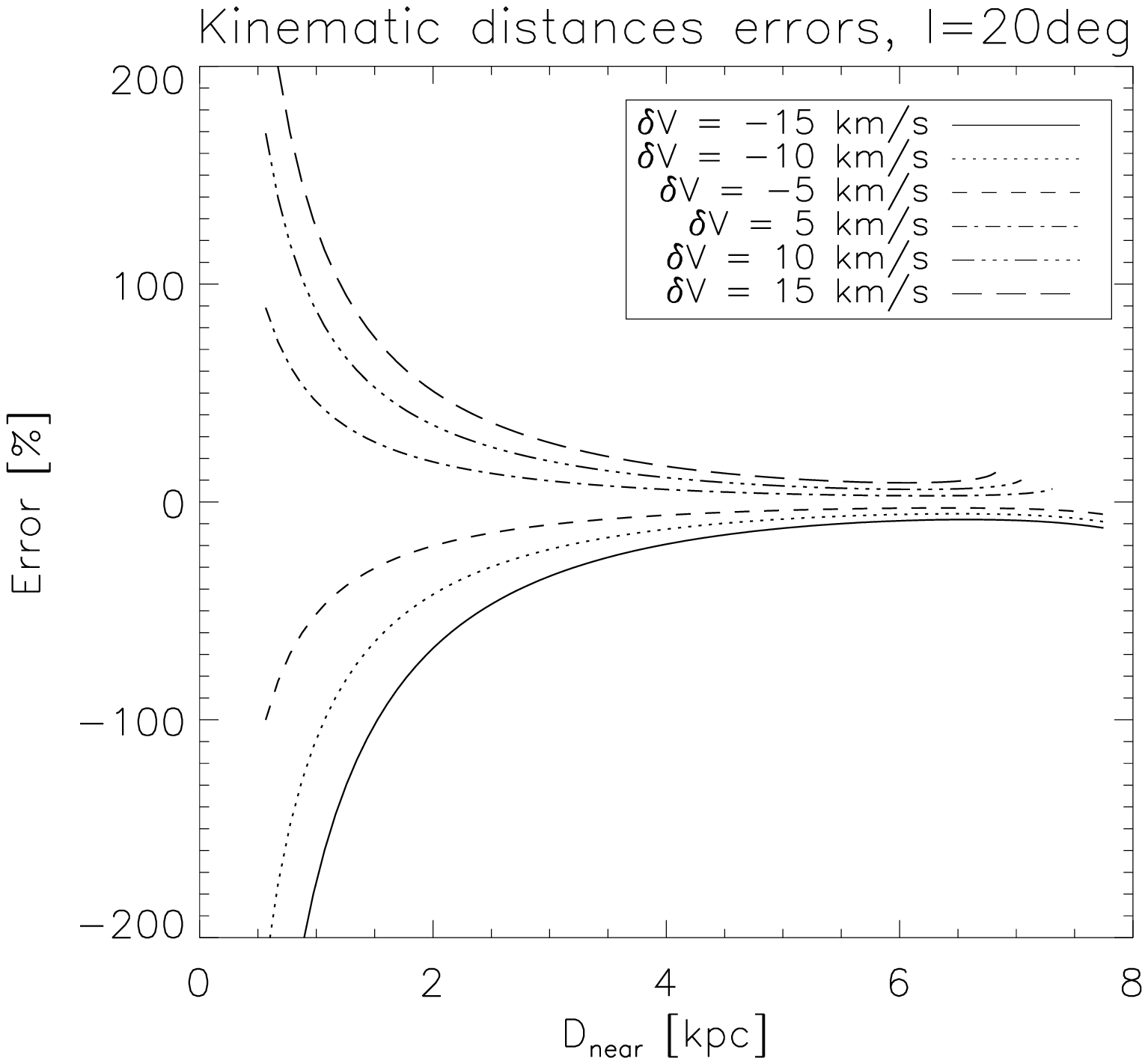}}
\subfigure{\includegraphics[height=7cm]{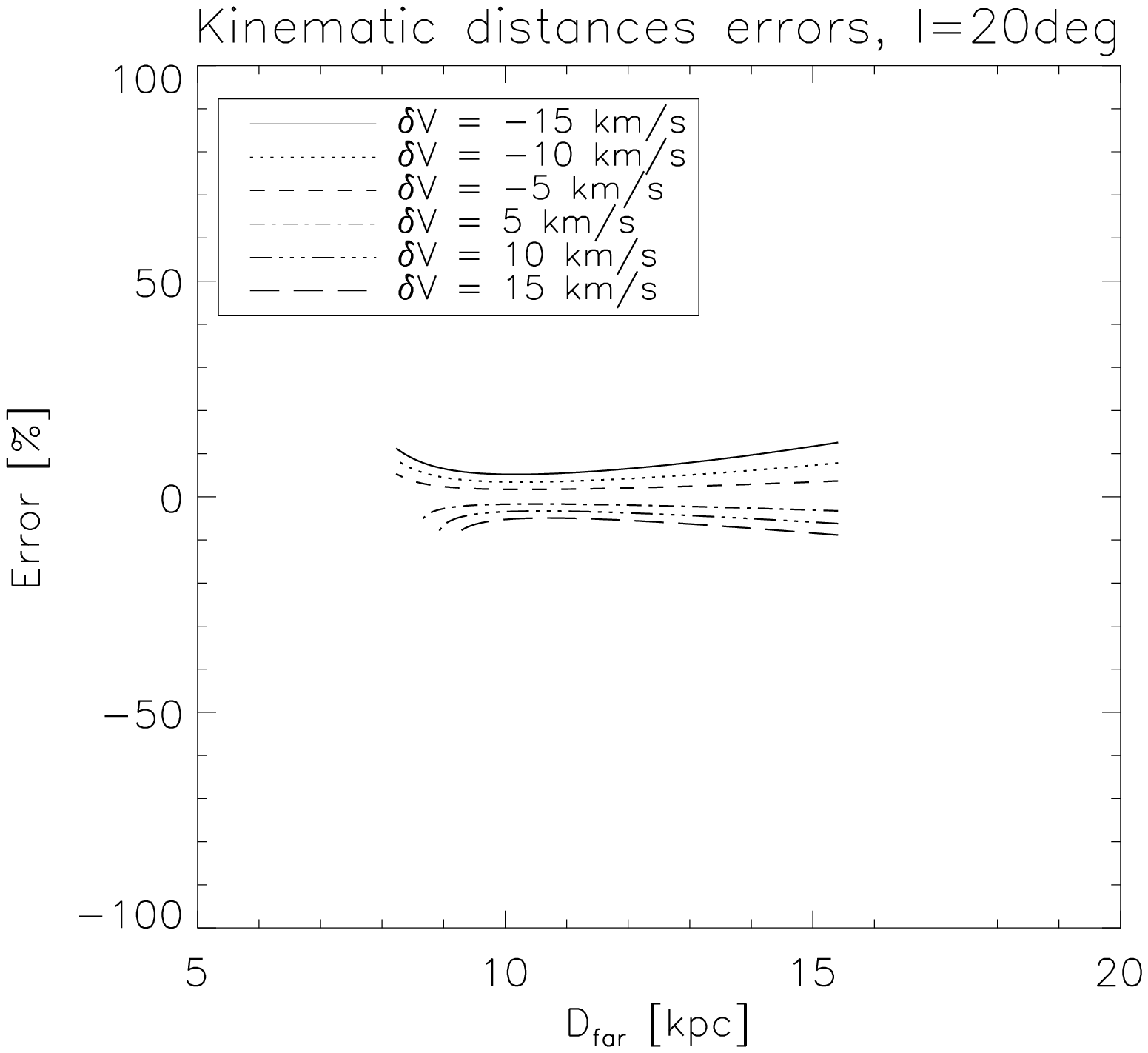}}
\subfigure{\includegraphics[height=7cm]{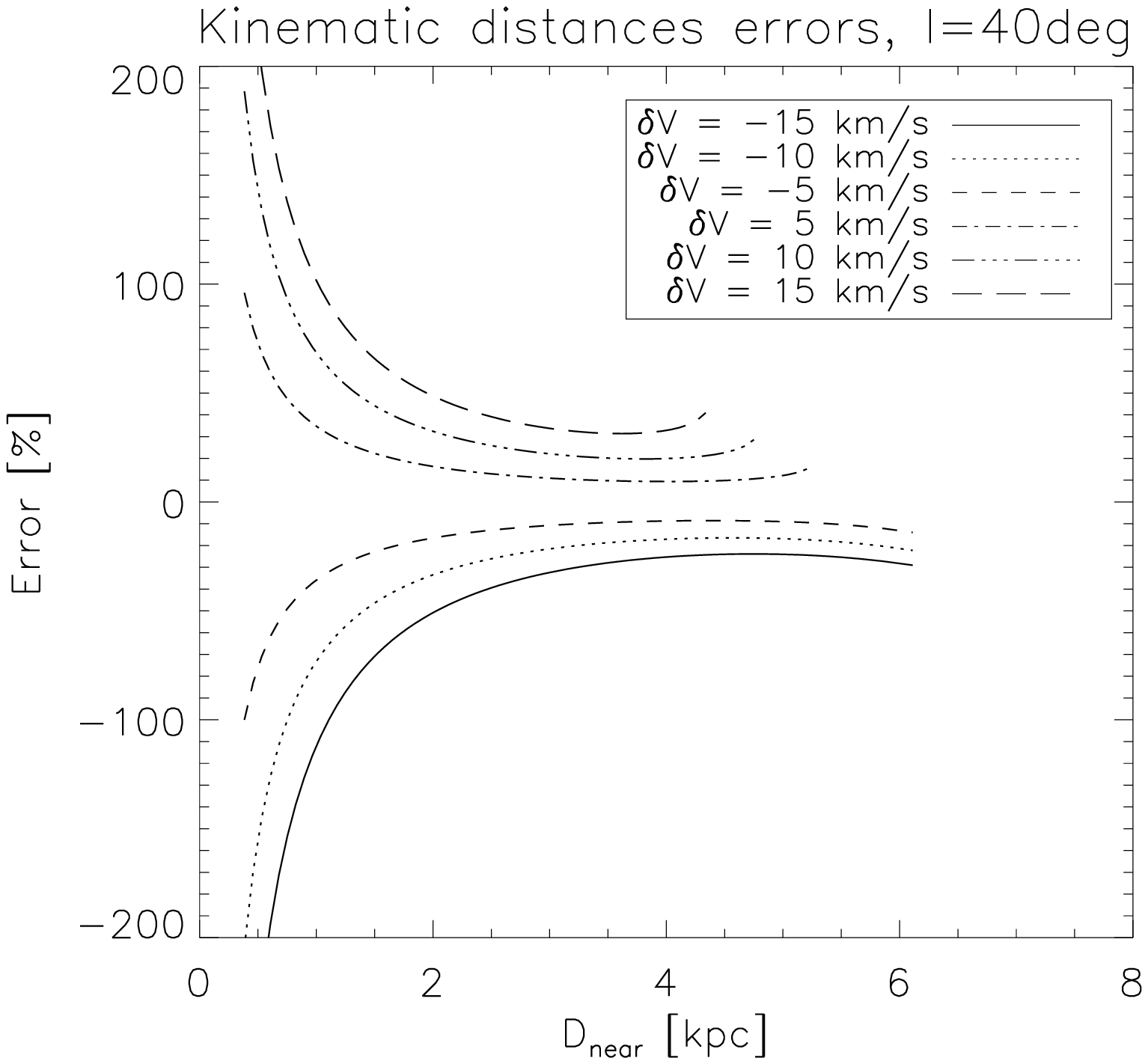}}
\subfigure{\includegraphics[height=7cm]{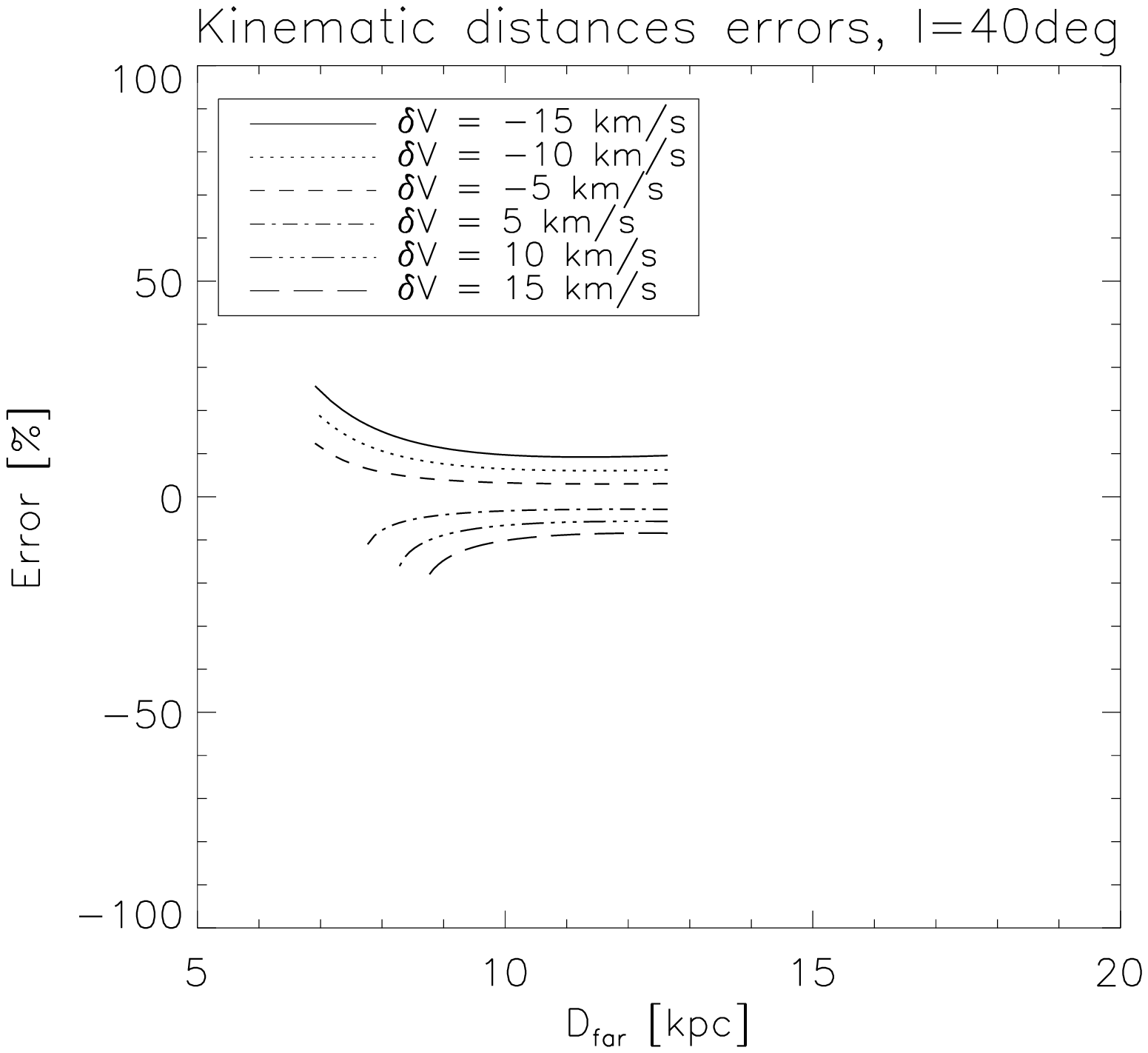}}
\caption{Error on the estimation of the near (left) and far (right) kinematic distances (from a flat rotation curve) for Galactic longitudes $\ell$ = 20\degree (top) and $\ell$ = 40\degree (bottom). The legend indicates the different errors $\delta$V commited on the radial velocity (and consequently LSR velocity) of the cloud because of random motions and velocity jumps associated for instance with spiral arms. Each line represents the error on the distance for a given error on the velocity.}
\label{err_dist}
\end{figure}

\indent There are several sources of errors in the estimation of the clouds' distances. Two are due to an error in the estimation of the LSR velocity of the cloud. We measure the radial velocity of clouds, or in other words the projection of their velocity vector around the Galactic center onto the line of sight. However, there is  a cloud-to-cloud velocity dispersion relative to the rotation curve, on the order of 3 \kms \citep{C85}. This difference between the velocity of a cloud and its LSR velocity  leads to an error in the estimation of its kinematic distance. 
Furthermore, spiral arms induce shocks associated with non-circular velocity discontinuities and velocity gradients (both in magnitude and direction) in the gas flow, producing a systematic error in the LSR velocity of a cloud compared to pure circular orbital motions. This error includes effects of non-circular motions and variations in the magnitude of the circular velocity of a cloud. The gravitational perturbation induced by the overdensity in the spiral arm will also produce a velocity perturbation. The exact calculation of the magnitude of the velocity jump as the gas flows accross the spiral arm is beyond the scope of this paper. However, we estimate the typical velocity perturbations to be 10 \kms to 15 \kms  as measured by \citet{C85}, who observed regions exhibiting motions of up to 15 \kms relative to the best-fit rotation curve and were geometrically associated with the Scutum-Crux, Sagittarius, and Local spiral arms.\\
\indent  Finally, additional errors are related to how CLUMPFIND extracts clouds from the GRS data. It estimates the radial velocity of the cloud as the velocity of the peak voxel in the molecular cloud, which yields a possible error in the estimation of the centroid velocity of the cloud. However, this source of error is negligible compared to the cloud-to-cloud velocity dispersion and the systematic error related to perturbations within spiral arms (the line-width of the clouds is in the order of 1 to 5 \kmsn).\\
\indent Figure \ref{err_dist} shows, for Galactic longitudes of 20\degree and 40\degn, the errors on the estimated kinematic distance for a range of near and far distances and for several errors in the estimation of the LSR velocity of a cloud. The fractional error can be quite large for the near solution. However, for distances greater than 3 kpc, which is where 85$\%$ of the \CO luminosity within GRS clouds originates from and where 80$\%$ of the GRS clouds are located, the error on the kinematic distance is at most 30$\%$ for the near distance and less than 20$\%$ for the far distance, assuming an LSR velocity error of 15 \kmsn. It is interesing to note that \citet{C85} found that the velocity discrepancies associated with spiral arms were greater that the velocity of the best-fit rotation curve, which would place the clouds farther than they really are if the clouds lie at the far kinematic distance, or closer if they lie at the near distance.

\section{The Galactic distribution of GRS clouds}\label{map}
\begin{figure}
\includegraphics[height= 11cm]{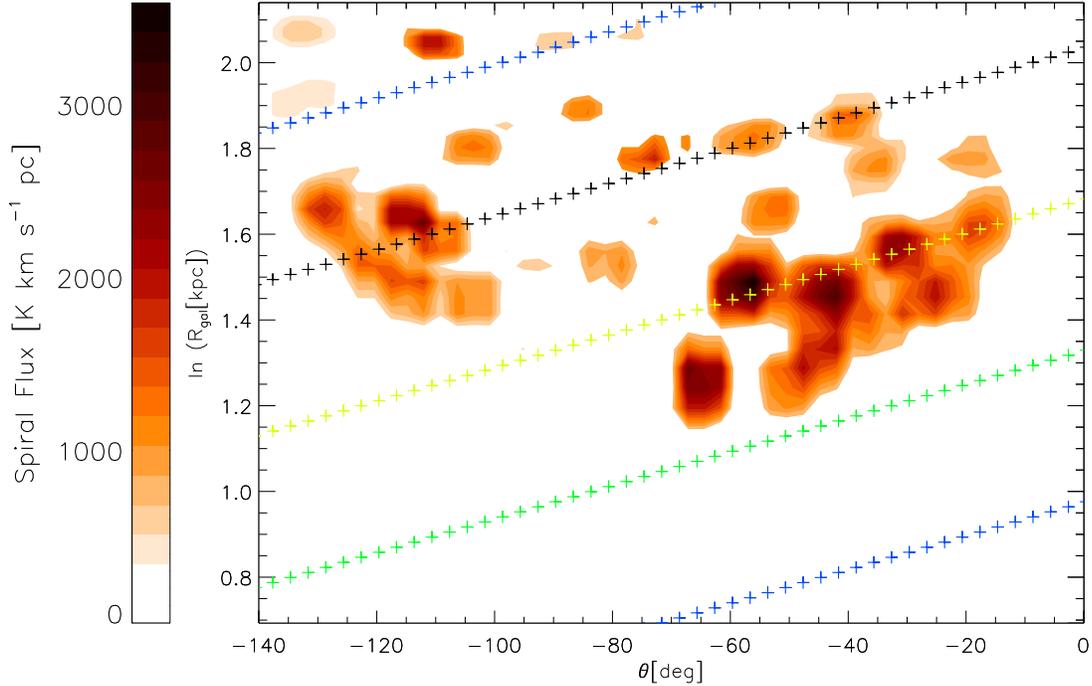}
\caption{\CO surface brightness from the GRS clouds plotted in ($\theta$, ln(r)) space (in units of K \kms pc). $\theta$ is the azimuth with the origin at the sun-GC axis and r is the galactocentric radius. In this space, spiral arms appear as straight lines. The loci of crosses represent the \citet{vallee95} four-arm model with a pitch angle of 12.7\degn. The green, yellow, black, and blue lines correspond to the 3 kpc-arm, the Scutum-Crux arm, the Sagittarius arm and the Perseus arm respectively. The surface brightness was obtained by summing the \CO luminosities of the GRS clouds located within each bin and dividing by the surface element, which in ($\theta$, ln(r)) space has units of pc. }
\label{plot_logr_theta_flux}
\end{figure}

\begin{figure}
\includegraphics[height= 6cm]{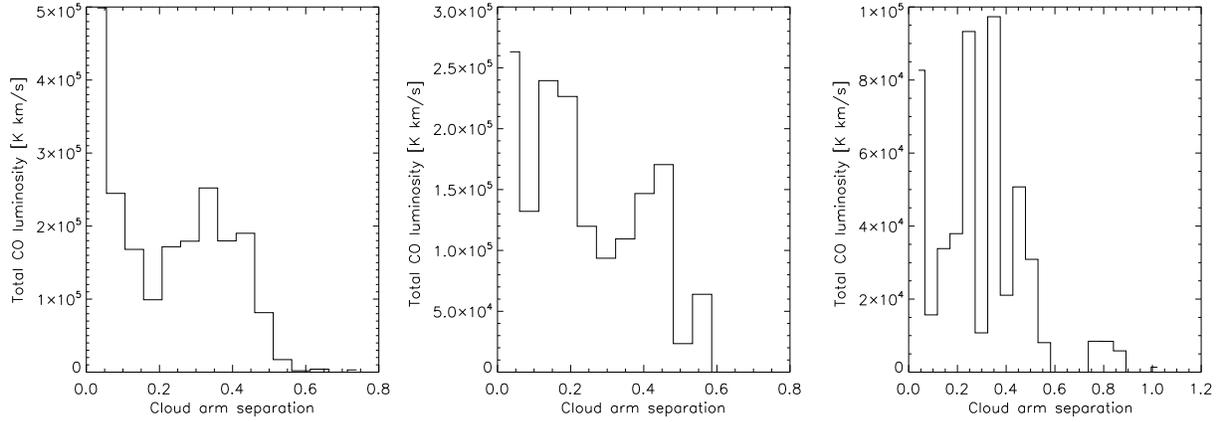}
\caption{\CO luminosity from GRS clouds as a function of cloud-to-spiral-arm separation, expressed as a fraction of the inter-arm separation. The left, middle, and right panels correspond to clouds for which the Scutum, Sagittarius, and Perseus arms respectively host the molecular clouds shown in the panel. The total \CO luminosity was computed by summing all the individual cloud \CO luminosities over each cloud-to-spiral-arm separation bin (0.05 $\times$ the inter-arm separation). 
}\label{spiral_arm_clouds}
\end{figure}

\indent The positions of the GRS clouds are shown in Figure \ref{map}. However, the locations and number density of clouds do not reflect the distribution of the molecular gas in the Galaxy because of resolution effects and the cloud identification algorithm. CLUMPFIND detects well resolved, very nearby clouds and clumps as distinct clouds while the same clouds at a farther distance would be assigned to a single cloud. It is indeed clear from the left panel of Figure \ref{map} that there are far more near clouds than far clouds detected by CLUMPFIND. Therefore, one needs to consider the Galactic surface mass density of molecular clouds in order to accurately represent the Galactic distribution of molecular gas. The \CO luminosity of a molecular cloud is a good proxy for its mass. Because \CO is optically thin compared to \COT, it probes H$_2$ column-densities and clouds' masses more accurately. Thus, the \CO Galactic surface brightness roughly reflects the distribution of the molecular mass content of the Milky Way. The right panel of Figure \ref{map} shows the Galactic \CO surface brightness from the GRS clouds obtained by summing the \CO luminosities of the clouds over bins of 0.04 kpc$^2$ and dividing by the bin surface. The resulting map was then smoothed by a box kernel of width 0.4 kpc.\\
\indent The distribution of molecular clouds in the Galaxy can be examined with respect to global models of the Milky Way based on different spiral arm tracers (e.g., star count, HII regions, OB associations, electron density, magnetic field). The Galactic distribution of molecular gas obtained with the GRS is compared to (1) a two-arm model based on the distribution of K and M-giants by \citet{benjamin08} and (2) to a four-arm model based on a compilation of tracers by \citet{vallee95}. The models are described below.\\
\indent If spiral arms are produced by a non-axisymetric component in the stellar distribution, which embraces the majority of the disk mass, then the pertubation itself should be best traced by the old stellar population using K band imagery \citep{drimmel00}, which constitutes an optimized balance between exctinction and contamination by hot dust. \citet{benjamin08} devised a two-arm model of the Milky Way based on GLIMPSE (Galactic Legacy Infrared Mid-Plane Extraordinaire) star count data and on the distribution of K and M-type giants. The surface density of the K giant population decreases significantly beyond a radius of $\sim$ 4.5 kpc from the Galactic center. The presence of a stellar bar with a length of $\sim$ 9 kpc and a major axis tilted by 44\degree with respect to the direction to the Galactic center is inferred from the variation of distances to these stars with respect to Galactic longitude.  The presence of the Galactic stellar bar has  also been deduced from IRAS, COBE and 2MASS data \citep{weinberg92, freudenreich98, cole02}. \citet{benjamin08} identified a peak in the star counts at longitudes 301\degree to 306\degree that indicates a tangency with a spiral arm. From these observations, they devised a two-arm model (Scutum-Crux and Perseus arms) of the Milky Way shown in the background image of Figure \ref{map}. \\
\indent \citet{vallee95} compiled models of the spiral structure of the Milky Way based on different tracers (e.g., CO emission, HII regions, HI emission, magnetic field, OB associations, and location of thermal electron gas) to produce a logarithmic four-arm model of our Galaxy (3 kpc, Scutum-Crux, Sagittarius, and Perseus arms). The \citet{vallee95} model is indicated by colored dashed lines in Figure \ref{map}. The distribution of the pitch angle obtained from different tracers is fairly wide. However, the statistical analysis performed in \citet{vallee95} yielded a mean pitch angle of 12.5\degn. The mean pitch angle obtained from previous molecular data \citep{blitz82, dame86, grab88, vallee95} was 12.7\degn, which is the value used in the model displayed in Figures \ref{map}. \\
\indent The comparison between the Galactic \CO surface brightness obtained from the GRS and models based on other tracers shows that the \CO surface brightness is strongly enhanced along the Scutum-Crux, Sagittarius and Perseus arms. This suggests that these spiral arms contain most of the molecular mass detected in the GRS. To further assess the spiral nature of the distribution of molecular gas,  Figure \ref{plot_logr_theta_flux} shows the \CO surface brightness map in ($\theta$, ln(r)) space, in which spiral arms appear as straight lines. $\theta$ is the azimuth with the origin located on the Galactic center-sun axis and r is the galactocentric radius. The surface brightness in this space has units of K \kms pc. Most of the \CO surface brightness is emitted from the Scutum-Crux arm. Although the Sagittarius and Perseus arms have a lower surface brightness, they also stand out as local extrema in the Galactic \CO surface brightness. \\
\indent The confinement of massive, \CO bright molecular clouds to spiral arms is further seen in Figure \ref{spiral_arm_clouds}, which shows the \CO luminosity contained in GRS clouds versus cloud-spiral-arm separation calculated as a fraction of the inter-arm separation. In the \citet{vallee95} model, the Scutum, Sagittarius, and Perseus arms have the following equations:

\begin{equation}\label{arm_eq}
r = 2.65 \mbox{ kpc } e^{tan(p)(\theta + \theta_0)}
\end{equation}

\noindent where $\theta$ is the azimuth around the galactic center (GC), with the origin on the sun-GC axis, $\theta_0$ = $\pi$, $3\pi/2$, and $2\pi$ for the Scutum-Crux, Sagittarius, and Perseus arms respectively, and $p$ = 12.7\degree is the pitch angle. Knowing the clouds' distances, the cloud-to-spiral arm separation is computed for each cloud, and the closest spiral arms is assigned as the parent spiral arm. From Equation \ref{arm_eq}, the inter-arm separation is given by:

\begin{equation}
r = 2.65 \mbox{ kpc } e^{tan(p)(\theta + \theta_s)}\left(e^{\frac{\pi}{2} tan(p)} -1 \right)
\end{equation}

\noindent where $\theta_s$ = $\pi$ for the separation between the Scutum-Crux and the Sagittarius arms, and $\theta_s$ = $3\pi/2$ for the separation between the Sagittarius arm and the Perseus arm. For each cloud, the azimuth angle is computed and the corresponding cloud-to-spiral-arm separation can then be expressed as a fraction of the inter-arm separation. In Figure \ref{spiral_arm_clouds}, the individual cloud \CO luminosities are summed up in each cloud-spiral-arm separation bin (we used a bin size of 0.05 $\times$ the inter-arm separation) to produce the \CO luminosity as a function of cloud-spiral-arm separation expressed as a fraction of the inter-arm separation. The left, middle and right panels correspond to clouds associated with the Scutum, Sagittarius, and Perseus arms respectively. For the Scutum-Crux and Sagittarius arms, the \CO luminosity is clearly a decreasing function of the cloud-to-spiral-arm separation. This supports the previous observation that molecular clouds are confined to spiral arms. For the Perseus arm the trend is not as clear, and the center of the spiral arms indeed to be offset with respect to the peak \CO luminosity. This is also seen in Figure \ref{map}, where massive molecular cloud complexes are slightly offset from the Perseus arm. The \CO luminosity is nonetheless concentrated in a narrow band consistent with a spiral arm. The small cloud sample associated with the Perseus arm does not allow us to determine its structure accurately.  \\  
\indent Altogether, Figures \ref{map}, \ref{plot_logr_theta_flux}, and \ref{spiral_arm_clouds} show that most of the molecular emission detected in the GRS clouds is located along the Scutum-Crux, Sagittarius, and Perseus spiral arms. The galactic distribution of molecular clouds obtained from the GRS is consistent with a four-arm model of the Milky Way, in which molecular clouds are confined to spiral arms. The locations of the Scutum-Crux and Perseus arms devised from the GRS are in agreement with models based on the old stellar population \citep{benjamin08} and on other tracers \citep{vallee95}. \\
\indent The Sagittarius arm is clearly detected in molecular data, but not in the old stellar population \citep[][model]{benjamin08}. The old stellar population (e.g., K- and M-giants) traces the non-axisymetric mass perturbation associated with spiral arms. On the other hand, HII regions, OB assocations, CO, and electron density trace the response of the gas and subsequent star formation to this mass perturbation. According to \citet{drimmel00}, this could explain why only two spiral arms are detected in the old stellar populations by \citet{benjamin08} and \citet{drimmel00}, while the distribution of molecular gas, HII regions, and O stars support four-arm models. \\

\section {Discussion}\label{discussion}

\subsection{Implications for molecular cloud formation}
\indent Molecular clouds can essentially form through three mechanisms: by coagulation of small, diffuse cloudlets (``bottom up'' formation), by pure instability (``top down'' formation) \citep[e.g.,][]{kim02, kim03, mousch74}, or by hydrodynamical processes (i.e., compression in shocks, in over-densities created by turbulence, in converging streams) \citep[e.g.,][]{audit05, heitsch05, vazquez06}. The latter models support the idea that molecular clouds form in spiral arm shocks \citep[e.g.,][]{blitz80, dobbs06, dobbs08, tasker09}, where warm atomic gas can be compressed and cooled via metal fine structure lines (mainly CII, O, SiII). The energy dissipation inherent to the shock, radiative cooling in the clumps, and orbit crowding then lead to the formation of molecular clouds within a few My \citep{bergin04}. \citet{blitz80} provided arguments against the coagulation model based on molecular cloud formation timescale. They showed that the formation of clouds by coagulation would require 10$^8$ years, and that the velocity dispersions predicted by coagulation models far exceeded the observed line-widths of molecular clouds. As for the instability models, \citet{kim02} investigated the role of swing amplification, Magneto-Jeans instability (MJI), and Parker instability in the formation of molecular clouds in sheared disks. They found that the MJI and swing amplification could cause condensations of a local Jeans mass to grow very rapidly. Similarly, \citet{kim03} showed that the Magneto-Rotational Instability (MRI) combined with swing amplification could lead to the formation of bound molecular clouds. \\
\indent The lifetime of molecular clouds is contrained by star formation activity. \citet{blitz80} showed that molecular clouds would be disrupted by the formation of OB associations within 10$^7$ years. However, they lacked the observational data to confirm their findings. \\
\indent Coagulation and instability models predict that molecular clouds form uniformly throughout the Galactic disk. In contrast, models that involve spiral shocks to form molecular clouds predict that they should be confined to spiral arms if they are short-lived as predicted by \citet{blitz80} (i.e., their lifetime is shorter than a traversal time between spiral arms). Assuming an inter-arm distance $\Delta$r = 0.3$r$ (obtained from the logarithmic \citet{vallee95} model), a global pattern speed of $\Omega_{gp}$ $=$ 11 \kms kpc$^{-1}$ \citep{lin67}, and a flat rotation curve V$_0$ = 220 \kmsn, the traversal time between the Scutum and Sagittarius arms amounts to 7 and 10 Myr at galactocentric radii of 5 kpc and 8 kpc respectively. Thus, the large-scale galactic distribution of molecular clouds can help constrain the lifetime of molecular clouds and differentiate between these different formation scenarios. Because the GRS covers a wide range of Galactic longitude, is fully sampled, and uses an relatively optically thin tracer ($\tau$ $\simeq$ 0.5 - 1), the \CO galactic surface brightness probes the mass distribution of molecular gas. The effects of \CO opacity on the relation between \CO luminosity and molecular column-density and mass are discussed in Section \ref{discussion}. Figures \ref{map}, \ref{plot_logr_theta_flux}, and \ref{spiral_arm_clouds} indicate that  molecular clouds detected in the GRS lie preferentially along spiral arms identified with other tracers. The galactic distribution of molecular clouds obtained from the GRS therefore supports cloud formation models that involve a spiral structure. In addition, the confinement of massive molecular clouds to spiral arms indicates that molecular clouds' lifetimes cannot exceed a traversal time between spiral arms (of order 10$^7$ years). If molecular clouds had a longer lifetime ($>$ a few 10$^7$ years), then the inter-arm space would be filled with massive, \CO bright molecular clouds that are not forming stars, which is not what we observe. This is further supported by the fact that the \CO luminosity (and hence molecular gas mass) is steeply decreasing with the distance to the parent spiral arm.  \\
\indent Not all the \CO flux detected in the GRS is assigned to molecular clouds by the cloud identification algorithm, due to the brightness level that defines the edge of a cloud being lower than the threshold applied to CLUMPFIND. Indeed, molecular clouds detected in the GRS account for 63 $\%$ of the total \CO luminosity in the GRS (i.e., CLUMPFIND recovers 63 $\%$ of the total GRS emission). The more diffuse \CO emission that is not emcompassed in the identified molecular clouds (i.e., 37 $\%$ of the GRS emission) originates from all locations in the Galaxy. Since this diffuse gas is not identified as clouds, the kinematic distance method and HISA cannot be used to estimate its location in the Milky Way. However, the very fact that, according to our observations, the inter-arm space can only be filled with diffuse molecular gas while massive, \CO bright molecular clouds are confined to spiral arms represents in itself a significant constraint in favor of cloud formation models involving spiral shocks.\\

\subsection {The choice of rotation curve}

\begin{figure}
\includegraphics[height= 10cm]{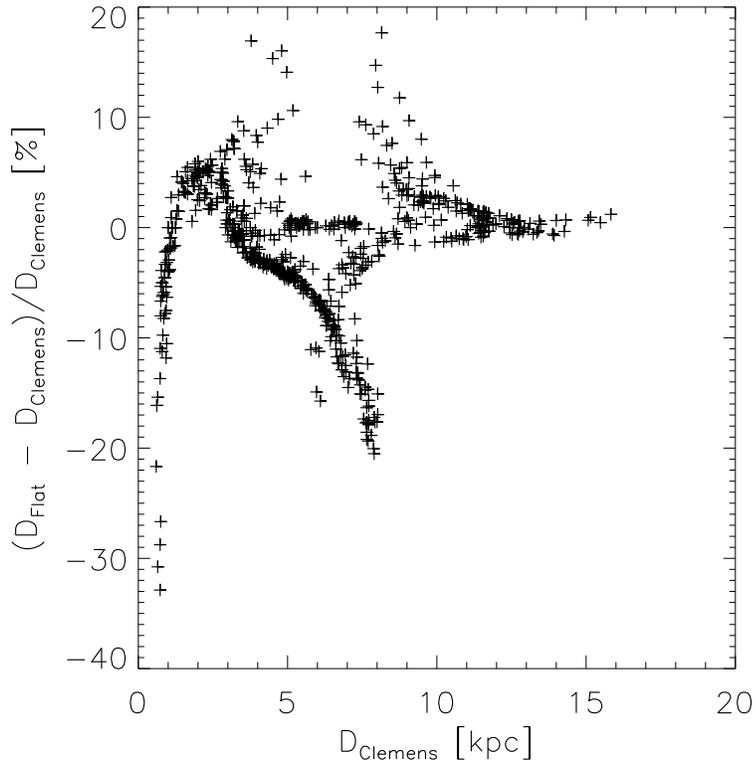}
\caption{Comparison between kinematic distances derived from the \citet{C85} rotation curve and a flat rotation curve with V$_0$ = 220 \kmsn. }
\label{compare_rot_curv}
\end{figure}

\indent We used the Clemens rotation curve in order to convert the radial velocity of a molecular cloud to a kinematic distance. This curve is derived directly from gas kinematics measurements (mostly CO, but also HI for $\ell$ $<$ 15\degn) at the tangent point (i.e where the radial velocity is maximum). To ascertain that our face-on map of the molecular clouds is insensitive to the exact choice of rotation curve, we repeated our analysis with a flat rotation curve of velocity V$_0$ = 220 \kmsn. This choice was motivated by the fact that flat rotation curves are observed in most galaxies, and that the Galactic circular velocity of 220 \kms is well established in the vicinity of the sun \citep[e.g.,][]{kerr86}. Furthermore, only in the very inner regions (r $< $ 2 kpc) will the rotation curve decrease toward 0 because of solid body rotation, but due to the limited longitude range of the GRS, the minimum galactocentric radius for all the GRS clouds is 2.2 kpc. Figure \ref{compare_rot_curv} shows the relative disagreement between distances derived from both rotation curves. The disagreement does not exceed 10$\%$ for most of the clouds, except at the tangent point where the difference can reach up to 30 $\%$. The significant disagreement between both distances at the tangent point arises from the fact that $\frac{\partial V_{LSR}}{\partial {D}} = 0 $ at the tangent point, where a small difference in velocity introduced by the difference between the two rotation curves will cause a large discrepancy in distance. Therefore, only cloud distances at the tangent point (i.e. at a distance of about 7.3 kpc for longitudes $\ell$ $\sim$ 15\degree and 5.5 kpc for higher longitudes $\ell$ $\sim$ 50\degn) will be affected by the choice of rotation curve. We computed the Galactic \CO surface brightness from GRS molecular clouds with distances derived from a flat rotation curve scaled to V$_0$ = 220 \kms (see the bottom panel of Figure \ref{map}). The distribution of molecular gas near the Galactic bar and the tangent point changes slightly, but the overall spiral structure of the Milky Way remains identical regardless of the rotation curve used to derive distances.

\subsection{Comparison with trigonometric parallax of masers}

\begin{figure}
\subfigure{\includegraphics[height=6cm]{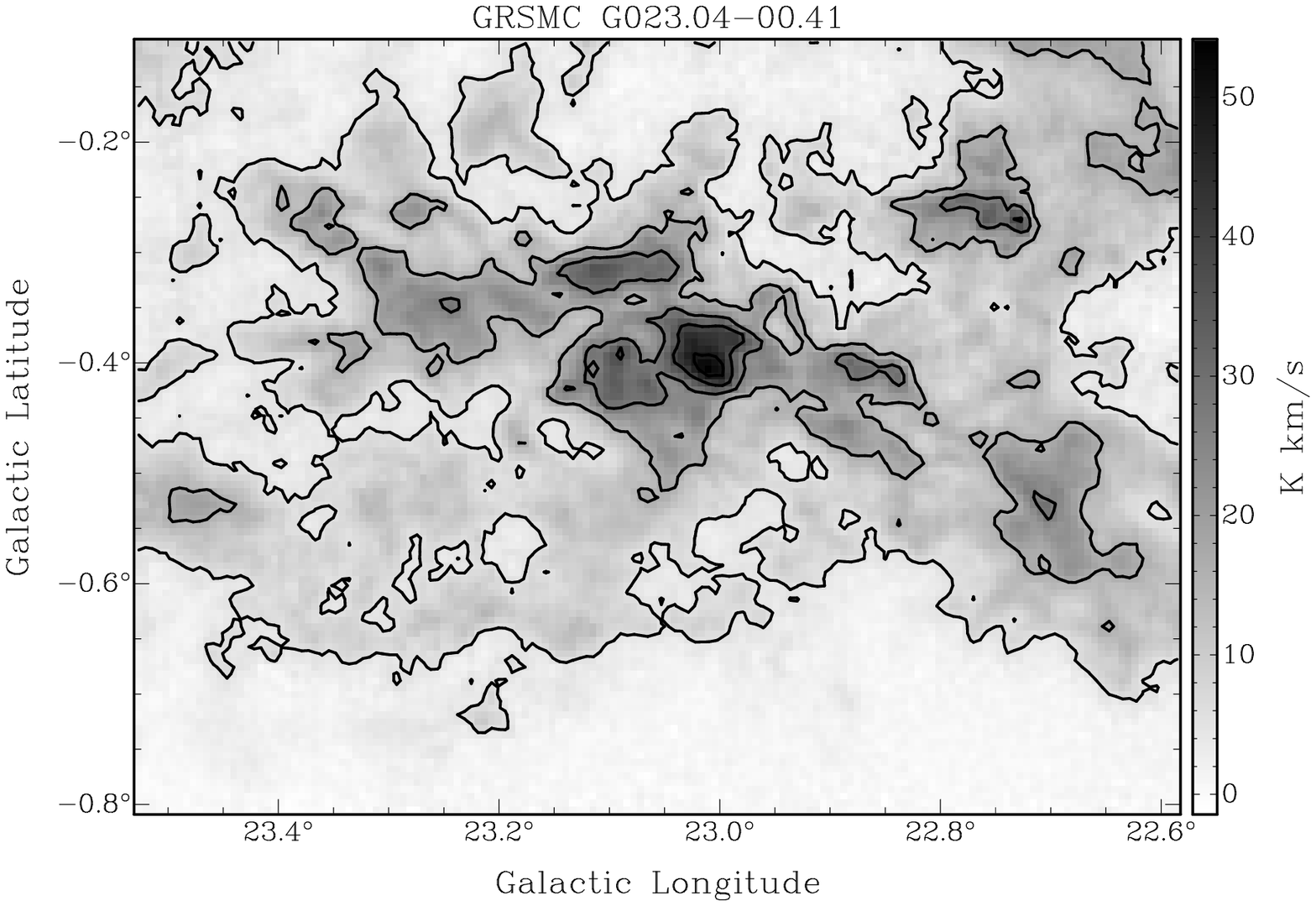}}
\subfigure{\includegraphics[height=6.5cm]{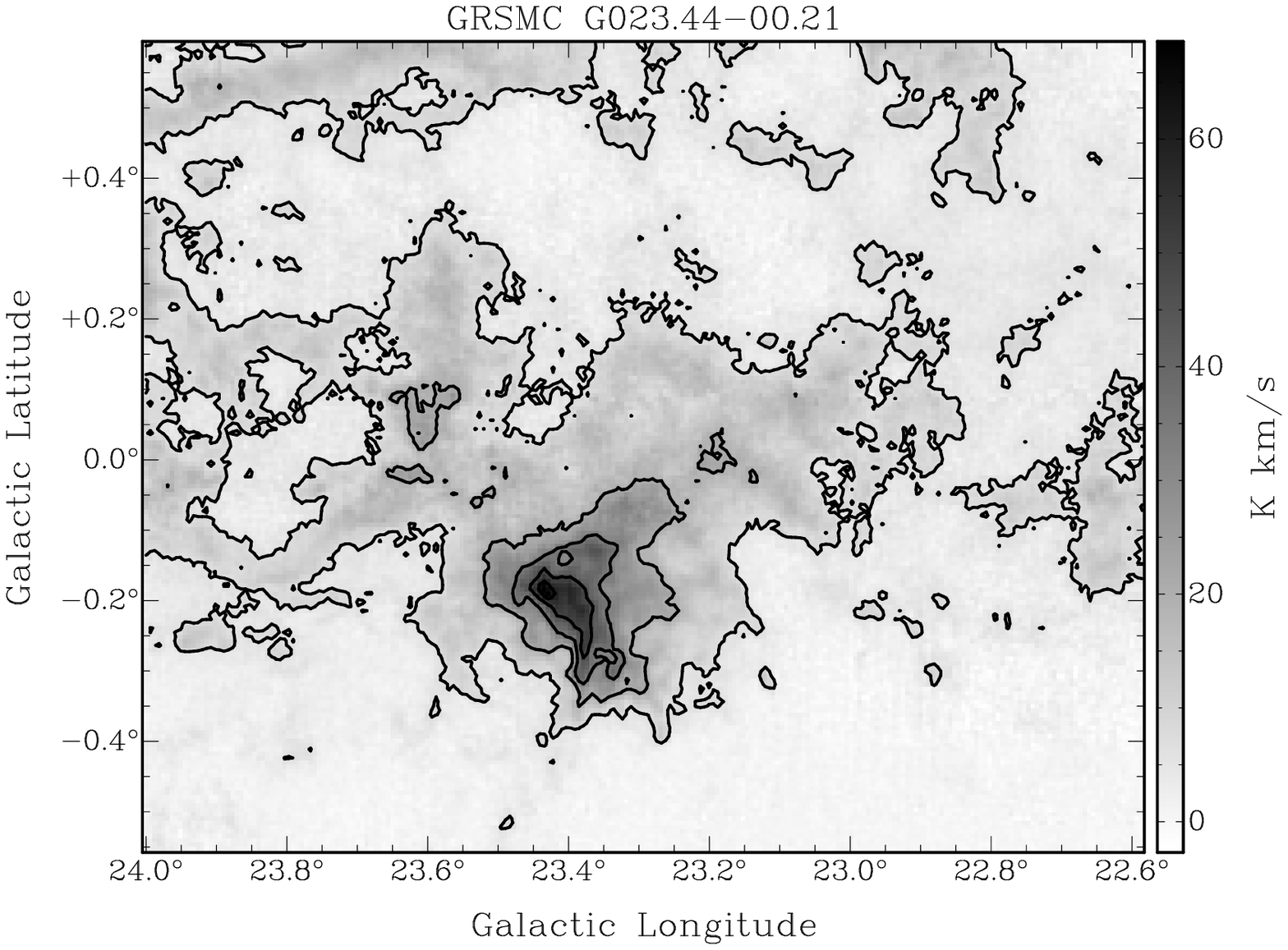}}
\subfigure{\includegraphics[height=6cm]{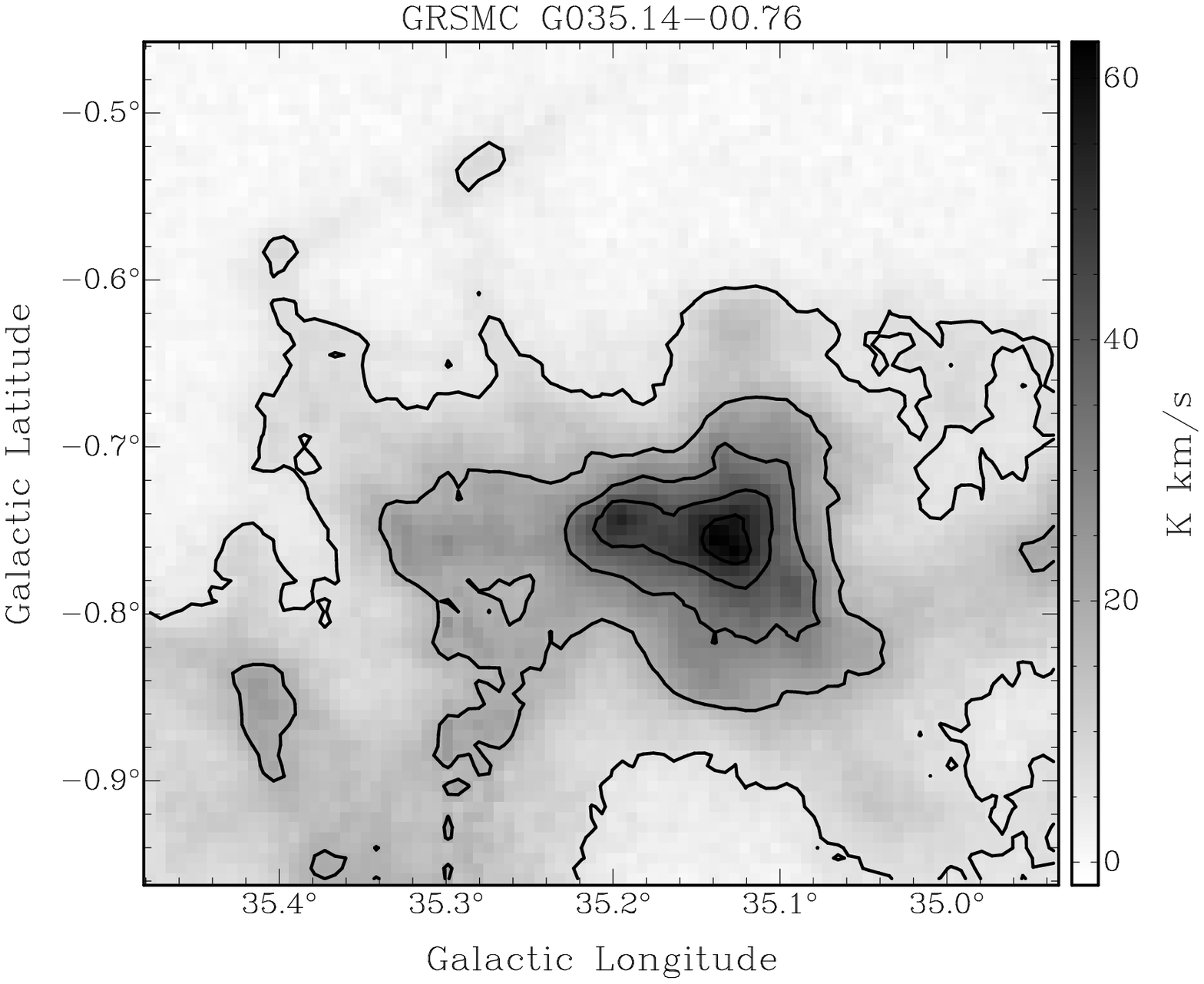}}
\subfigure{\includegraphics[height=6.5cm]{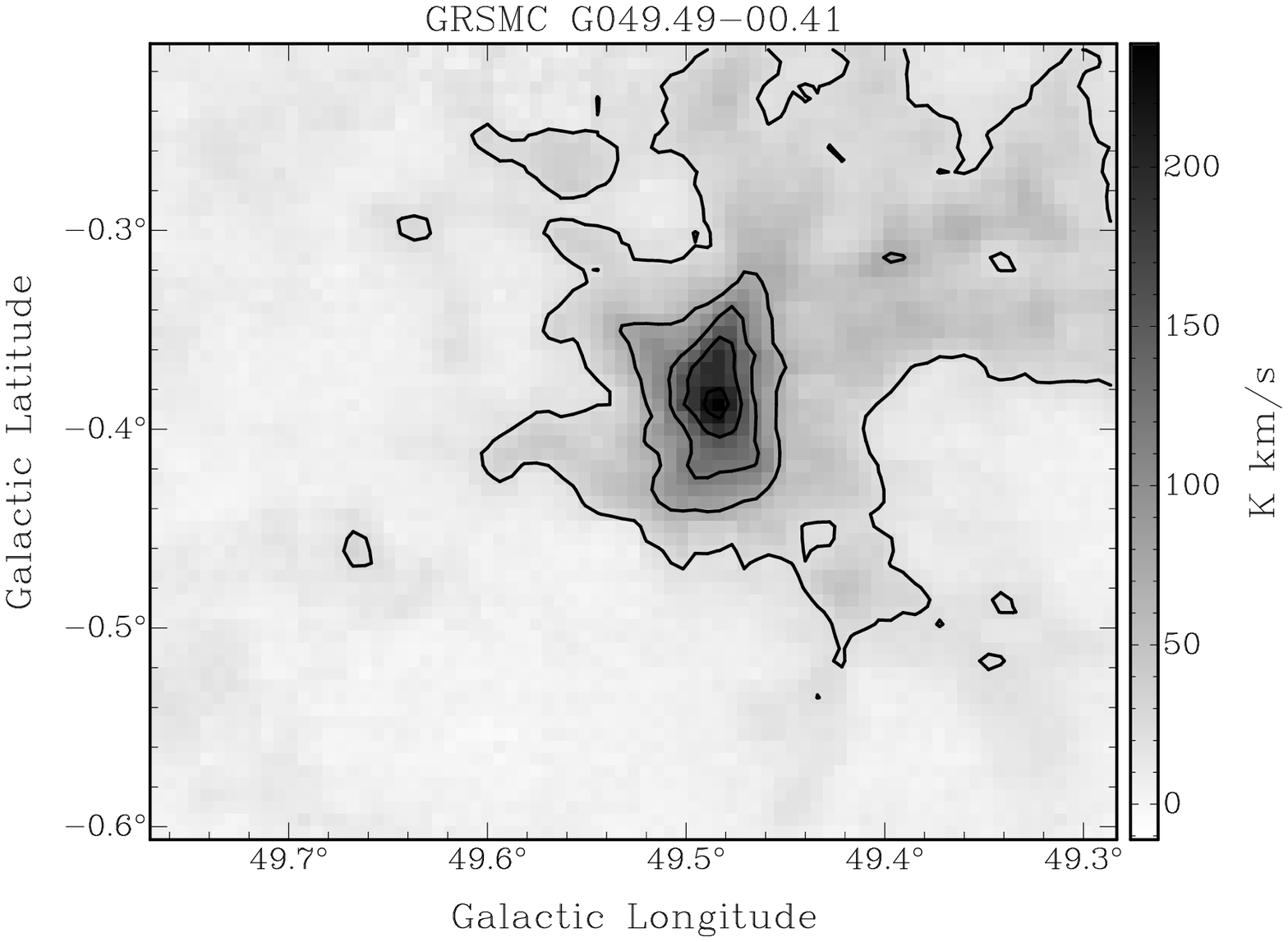}}
\caption{\CO integrated intensity images for four molecular clouds (GRSMC G023.04-00.41, GRSMC G023.44-00.21, GRSMC G035.2-00.74, and GRSMC G049.49-00.40 (W51 IRS2) respectively from left to right and top to bottom) for which distances from maser trigonometric parallax are available. The contour levels are 10\%, 30\%, 50\%, 70\% and 90\% of the peak integrated intensity (54 K \kmsn, 68 K \kmsn, 63 K \kmsn, and 238 K \kms for these four clouds respectively).}
\label{reid_co}
\end{figure}

\begin{figure}
\subfigure{\includegraphics[height=7cm]{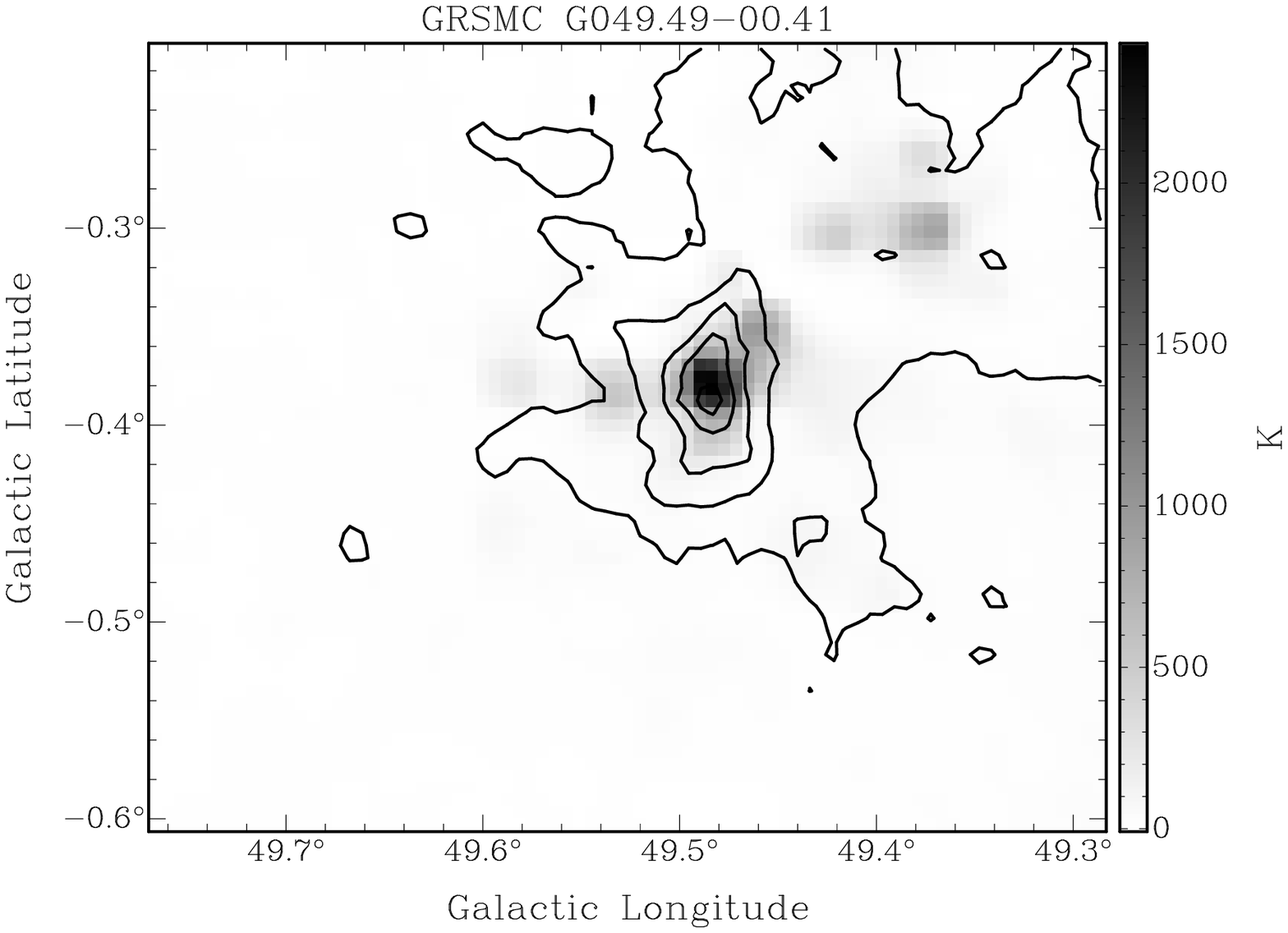}}
\subfigure{\includegraphics[height=7cm]{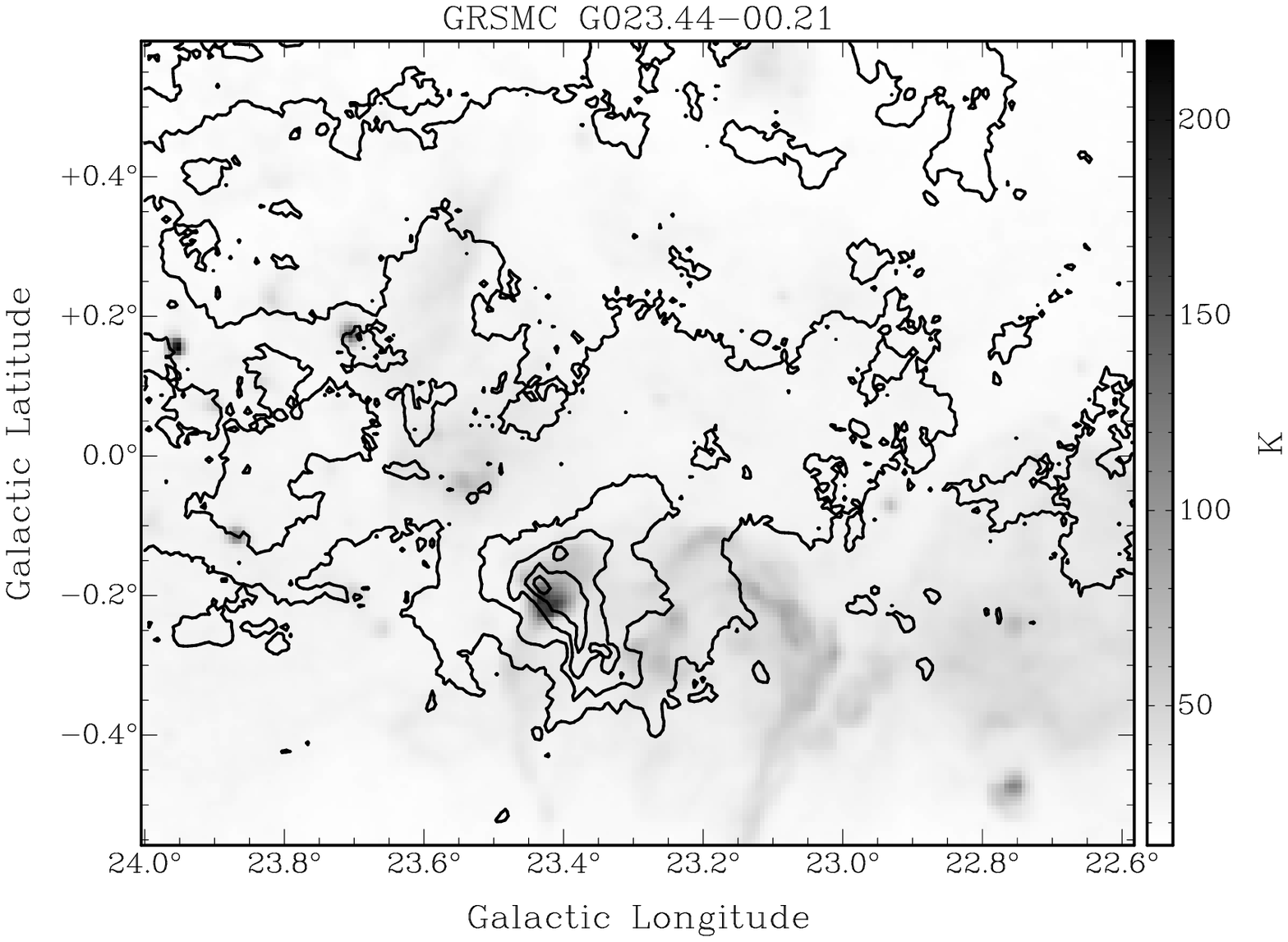}}
\caption{21 cm continuum images of GRSMC G049.49-00.40 (W51 IRS2) (left) and GRSMC G023.44-00.21 (right), two molecular clouds for which distances derived from trigonometric parallax of masers are available. The contour levels are 10\%, 30\%, 50\%, 70\% and 90\% of the peak integrated intensity (238 K \kms and 68 K \kms respectively). In both cases, 21 cm continuum sources are coincident with the \CO emission peaks of the clouds. These two molecular clouds thus contain 21 cm continuum sources, which can be used to resolve the kinematic distance ambiguity.}
\label{cont_reid}
\end{figure}

\begin{figure}
\subfigure{\includegraphics[height=8.5cm]{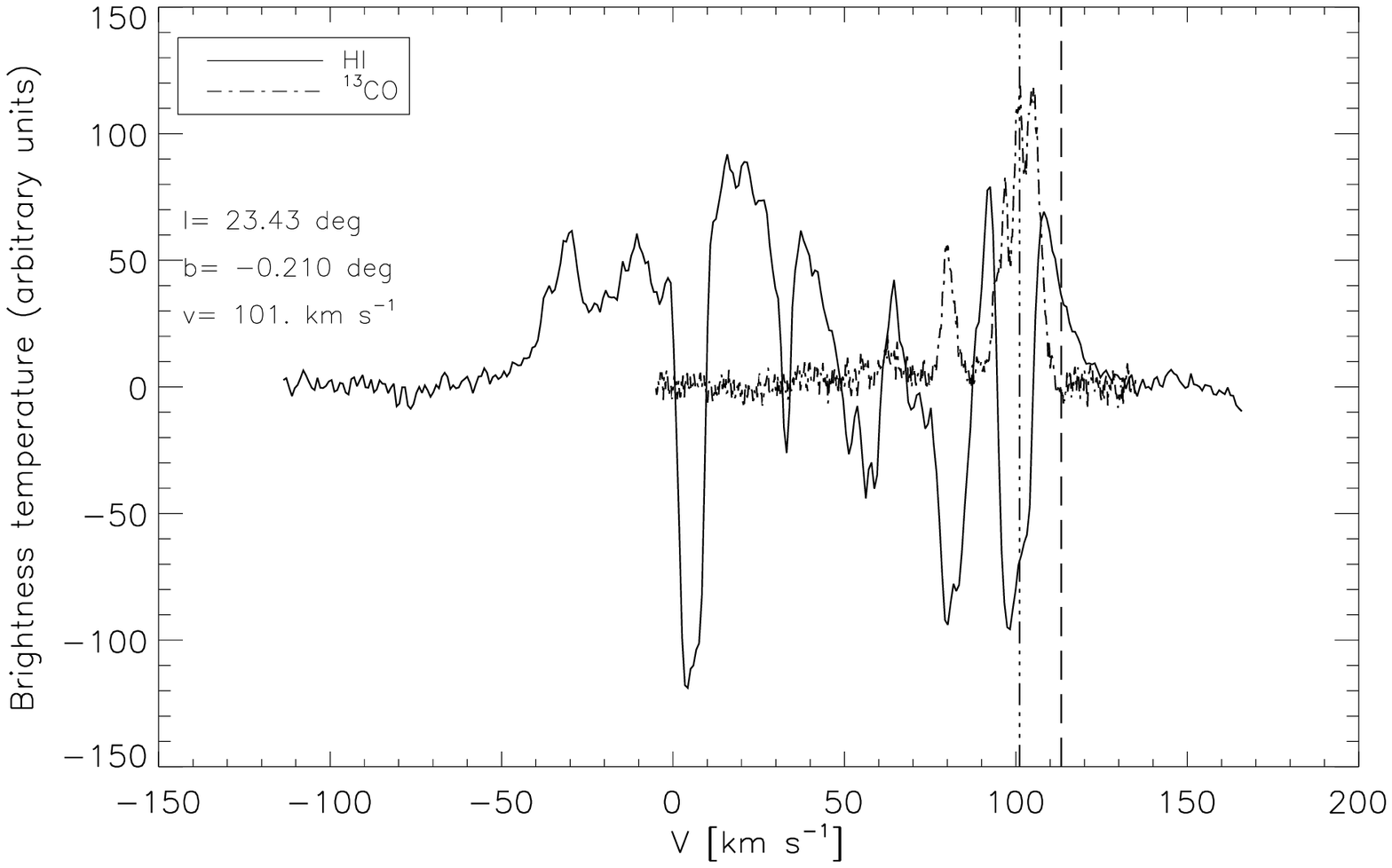}}
\vspace{0.05in}
\subfigure{\includegraphics[height=8.5cm]{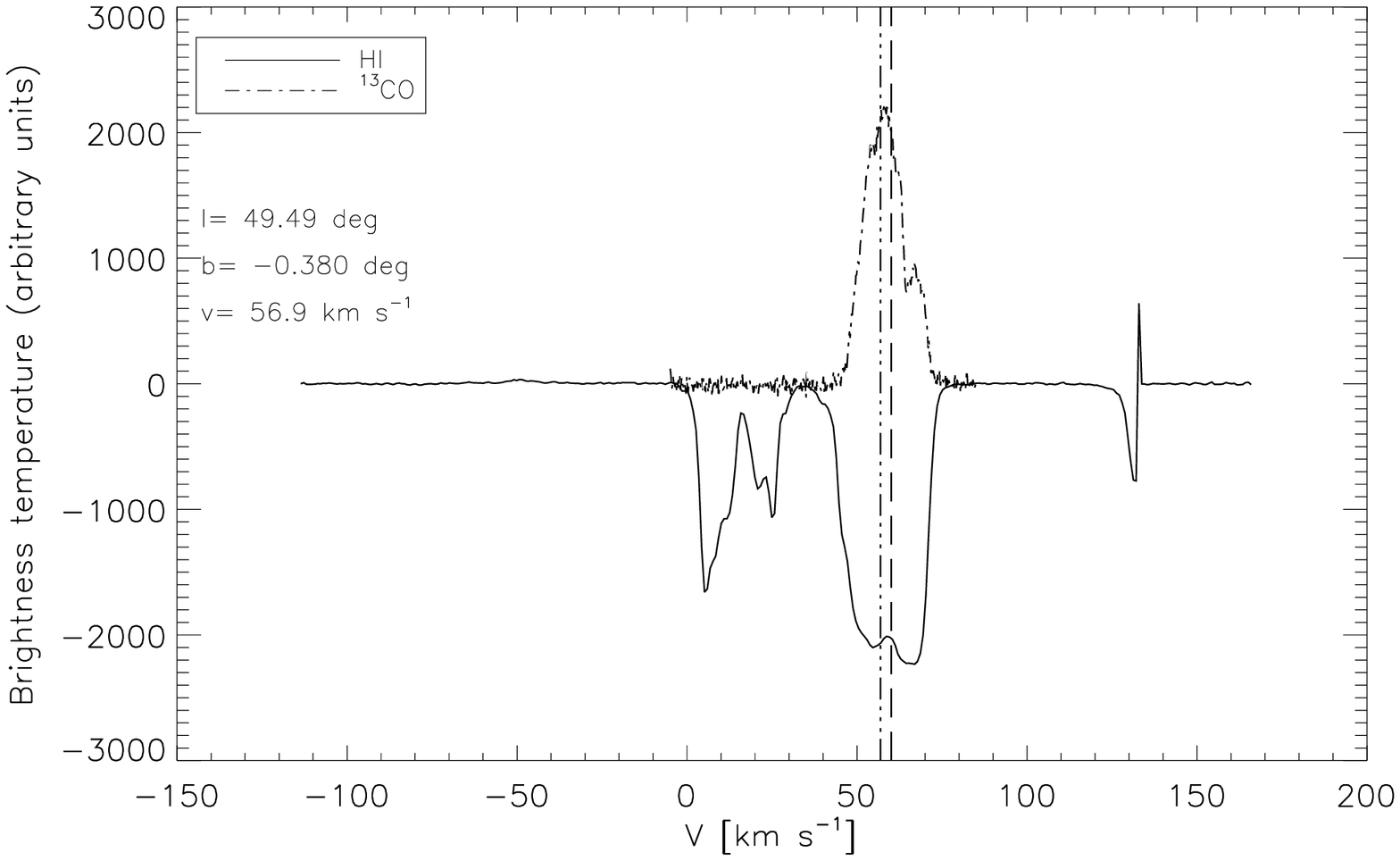}}
\caption{HI 21 cm spectra toward 21 cm continuum sources embedded in GRSMC G023.44-00.21 (top) and GRSMC G049.49-00.40 (W51 IRS2) (bottom), two molecular clouds for which distances derived from trigonometric parallax of masers are available. The dash-triple-dot line indicates the velocities of the clouds, while the dashed line indicates the velocity of the tangent point. In both cases, the 21 cm continuum is absorbed by foreground \CO bright molecular clouds up to the velocity of the clouds of interest, which indicates that these molecular clouds are located at the near kinematic distance (6.5 kpc and 5.5 kpc respectively). Note that in the case of W51 IRS2, there is a CO emission feature and corresponding absorption in the 21 cm continuum at a velocity 10 \kms greater than the velocity of the tangent point. This is most likely due to random, peculiar motions of molecular clouds with respect to their LSR velocity, which can cause the velocity of the gas to locally exceed the velocity of the tangent point.}
\label{cont_spec_reid}
\end{figure}

\begin{figure}
\subfigure{\includegraphics[height=8.5cm]{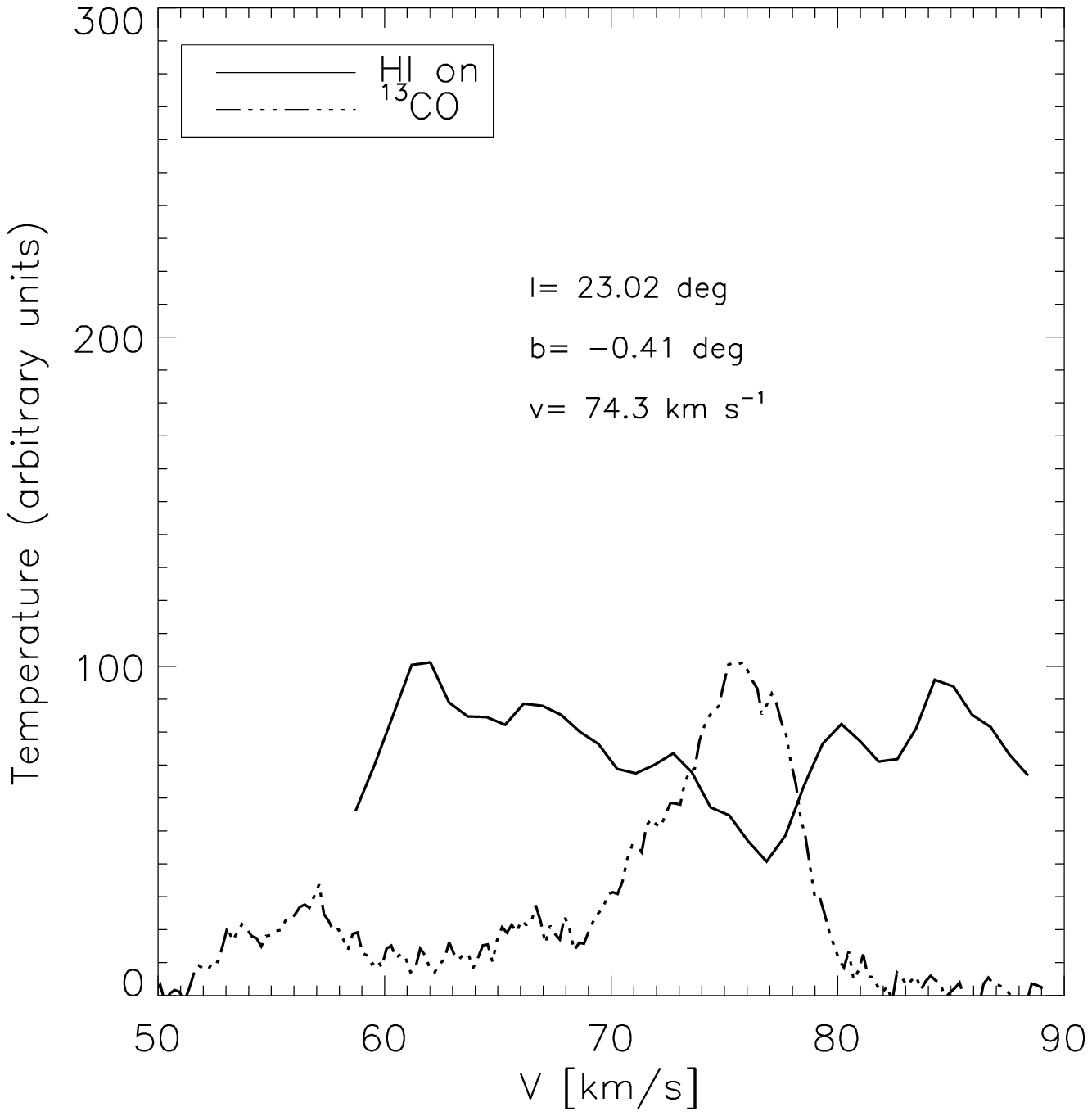}}
\subfigure{\includegraphics[height=8.5cm]{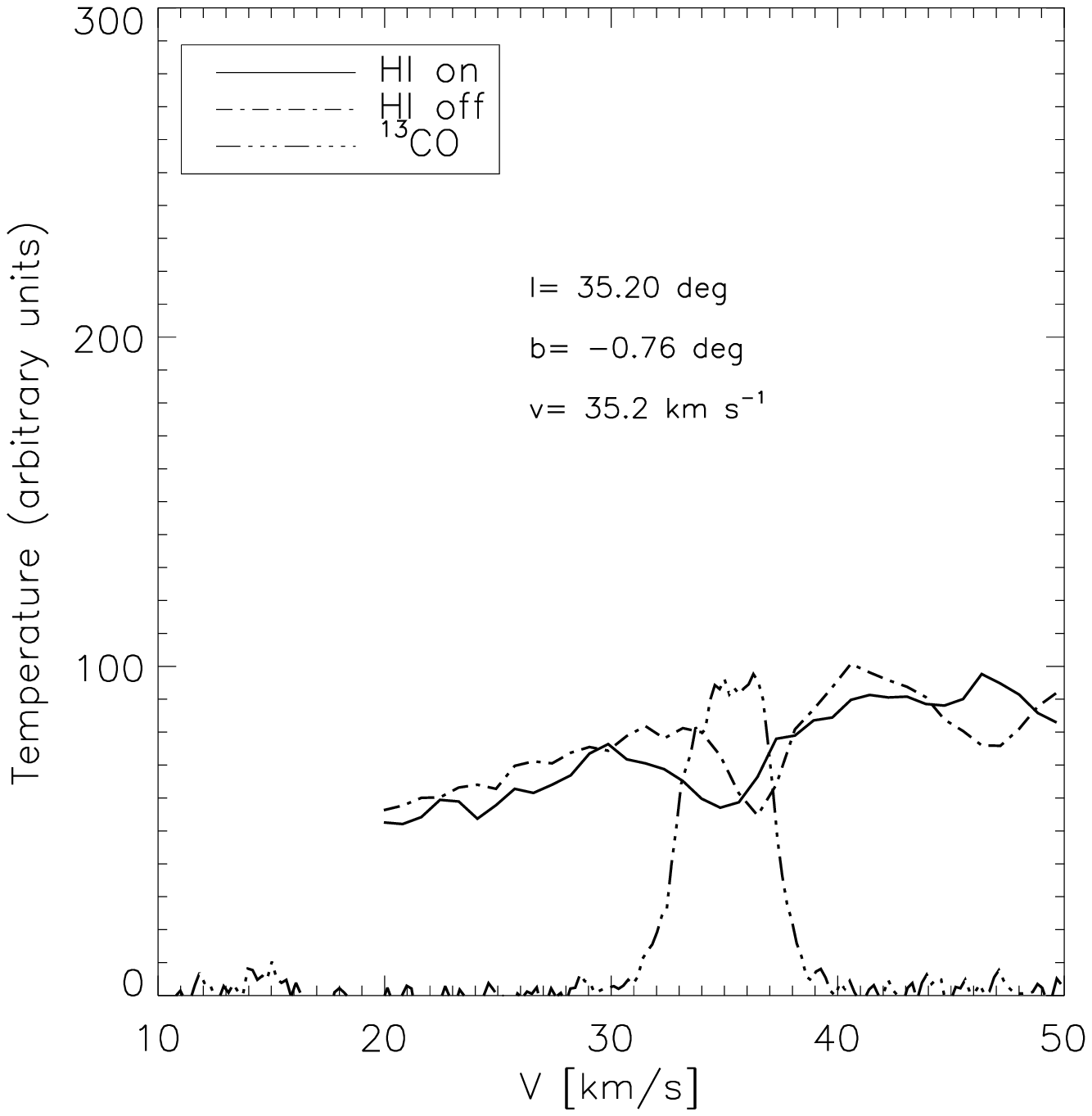}}
\caption{HI 21 cm ``on'' and ``off'' and \CO spectra toward GRSMC G023.04-00.41 (left) and GRSMC G035.2-00.74 (right), two molecular clouds for which distances derived from trigonometric parallax of masers are available. The absorption feature in the HI 21 cm ``on'' spectrum compared to the ``off'' spectrum at the velocity of \CO emission suggests that HISA is present toward these clouds. Therefore they are located at the near kinematic distance (4.9 kpc and 2.35 kpc respectively). }
\label{hisa_spec_reid}
\end{figure}

\indent Local velocity perturbations associated with spiral arms, expanding shells and non-circular motions near the Galactic bar affect the estimation of kinematic distances. For a perturbation of 15 \kmsn, Figure \ref{err_dist} shows that the error on the near kinematic distance can reach 50$\%$ for distances lower than 2 kpc. To provide accurate distances that do not rely on a kinematic model of the Galaxy, other authors have used sources embedded in molecular clouds rather than the clouds themselves to derive their distances. Trigonometric parallaxes of masers have yielded  distances accurate to 10$\%$ to a handful of sources \citep[e.g.,][]{brunthaler08, xu06, xu08, zhang08}. The advantage of this technique is that it does not rely on any assumption about the kinematics of the Galaxy involved. The method is thus a powerful tool to estimate the errors commited on the kinematic distance, and to assess the accuracy of a given rotation curve. Despite their accuracy, trigonometric parallax of masers could not be implemented on a large sample of molecular clouds because it requires large observing timelines (typically two years). Moreover, not all molecular clouds contain masers limiting the usefullness of the technique to only a few clouds. In this section, we compare kinematic distances and distances obtained from maser trigonometric parallax for 4 molecular clouds identified in the GRS.   \\
\indent In the past year, \citet{brunthaler08}, \citet{zhang08} and \citet{xu08} derived distances to star forming regions using the trigonometric parallax of methanol masers. They used the Very Long Baseline Array (VLBA) to determine the trigonometric parallax of strong methanol (CH$_3$OH) masers at 12 GHz. The goal was to study the spiral structure and kinematics of the Milky Way. Four of those masers fall in the range of the GRS and are located at positions ($\ell$ = 35.2\degn, $b$ = -0.74\degn, $v$ = 31 \kmsn), ($\ell$ = 23.01\degn, $b$ = -0.41\degn, $v$ = 74.32 \kmsn), ($\ell$ = 23.44\degn, $b$ = -0.21\degn, $v$ = 101.1 \kmsn) and ($\ell$ = 49.49\degn, $b$ = -0.37\degn, $v$ = 56.9 \kmsn). The \CO counterparts of these star forming regions were identified in the GRS as clouds by CLUMPFIND at very close positions (the brightest \CO voxel was located within 2' of the position of the maser). Figure \ref{reid_co} shows \CO integrated intensity images for these four molecular clouds. GRSMC G023.44-00.21 and GRSMC G049.49-00.40 (W51 IRS2) contain 21 cm continuum sources as shown in Figure \ref{cont_reid}. In Figure \ref{cont_reid}, bright continuum sources are coincident with the \CO emission peaks of GRSMC G023.44-00.21 and GRSMC G049.49-00.40. Therefore, the kinematic distance ambiguity was resolved using absorption features in the 21 cm continuum in both cases. Figure \ref{cont_spec_reid} shows that the 21 cm continuum toward these two clouds is absorbed up to their respective radial velocities, but not up to the velocity of the tangent point. Thus, they are both located at the near kinematic distance (6.5 kpc and 5.5 kpc respectively). For GRSMC G023.04-00.41 and GRSMC G035.2-00.74, the kinematic distance ambiguity was resolved using HISA. Figure \ref{hisa_spec_reid} shows that absorption features in the HI 21 cm spectra toward GRSMC G023.04-00.41 and GRSMC G035.2-00.74 are coincident with the \CO emission lines from these two clouds. Thus, HISA is present toward GRSMC G023.04-00.41 and GRSMC G035.2-00.74. These two clouds are therefore located at the near kinematic distance (4.9 kpc and 2.35 kpc respectively).\\
\indent  Table \ref{table_reid} summarizes the comparison between distances derived in this paper and distances derived from trigonometric parallax of methanol masers. The names of the molecular clouds are indicated in the first column, while columns 2 and 3 provide the positions of the masers. Column 4 indicates the radial velocities of the \CO emission peaks of the clouds, while column 5 provides the radial velocities of the masers. Distances derived from both methods are compared in columns 6 and 7. Kinematic and maser distances agree within 10 $\%$. Kinematic distances are therefore as accurate as distances obtained from maser trigonometric parallax for these particular clouds, indicating that the \citet{C85} rotation curve must be accurate in this region of the Galaxy.

\subsection{Effects of the \CO opacity and excitation temperature}

\indent In this paper, the \CO galactic surface brightness was used as a proxy for the mass distribution of molecular clouds. The high \CO surface brightness near the Scutum-Crux arm can however be explained by both a high mass molecular content and high temperatures within the clouds due to the more intense Galactic radiation field at low galactocentric radii \citep[e.g.,][]{mathis83, sodroski97}. \CO emission can also be optically thick in the densest parts of molecular clouds ($\tau(^{13}CO) \simeq 1)$ such that the \CO flux does not trace molecular column-densities accurately in these regions. These two effects might change slightly the distribution of molecular gas derived in Section \ref{map}, but they will not change the bulk of the galactic distribution of molecular clouds traced by the \CO surface brightness and luminosity. Estimates of the excitation temperatures and optical depths of molecular clouds are required in order to estimate the effects of \CO opacity on the relation between \CO luminosity and molecular mass and accurately derive the mass distribution of molecular gas from \CO observations. This can be done using \COT and \CO observations. Because \COT emission is saturated ($\tau(^{12}CO) > 50$), the \COT brightness temperature probes the CO excitation temperature of molecular clouds. The \CO optical depth and corresponding \CO and H$_2$ column-densities can then be derived from \CO observations. We have already computed the physical properties (mass, density, temperature) of molecular clouds following this method, and have found that the galactic mass distribution of molecular clouds is indeed well traced by the galactic \CO surface brightness. These results will be published shortly in a second paper, where we derive physical properties of molecular clouds detected in the GRS using both \CO and \COT data from the GRS and the University of Massachusetts - Stony Brook (UMSB) surveys to account for the \CO optical depth and excitation temperature of molecular clouds. 

\section{Conclusion}\label{conclusion}
\indent A sample of 829 molecular clouds was identified by \citet{rathborne08} in the Galactic Ring Survey, a \COJ  survey of the longitude range 18\degree $\leq$  $\ell$  $\leq$  55.7\degree and latitude range $-$1\degree $\leq$  $b$  $\leq$  1\degree of the Galactic plane. Kinematic distances to 750 of those clouds were derived using the \citet{C85} rotation curve. HI self-absorption and 21 cm continuum sources whenever available were used to resolve the kinematic distance ambiguity. The identified clouds, the \CO luminosity of which constitutes 63$\%$ of the total emission in the GRS, lie primarily along the Scutum-Crux, Sagittarius and Perseus spiral arms. The molecular gas distribution obtained from the GRS is consistent with a logarithmic four-arm model for the Milky Way from \citet{vallee95} based on the distributions of HII regions, CO, O stars and on the magnetic field and electron density. Only two arms are detected in the distribution of the old stellar population (the Scutum-Crux and Perseus arms), most probably because the old stellar population traces the non-axisymetric mass perturbation itself, while CO, HII regions, and O stars probe the response of the gas to the perturbation. However, the locations of the Scutum-Crux and Perseus arms obtained from the GRS are consistent with the K- and M-giant distribution obtained from GLIMPSE. The galactic distribution of molecular clouds obtained in this paper is strongly enhanced along spiral arms, while the inter-arm space is devoid of massive molecular clouds. This supports cloud formation mechanisms that involve a spiral arm structure rather than models based on pure instability or coagulation.  We also conclude that molecular clouds must be short-lived (with lifetimes $<$ 10$^7$ years) in order to explain the absence of massive, \CO bright molecular clouds in the inter-arm space.\\

\acknowledgments{
This work was supported by NSF grant AST-0507657. The molecular line data used in this paper is from the Boston University (BU)-FCRAO GRS, a joint project of Boston University and the Five College Radio Astronomy observatory funded by the National Science Fundation under grants AST 98-00334, AST 00-98562, AST 01-00793, AST 02-28993, and AST 05-07657. The authors also gratefully acknowledge the contribution of Robert Benjamin, who provided the two-arm model of the Milky Way, and the referee, whose careful reading resulted in a much improved presentation. }

{}

\clearpage
\begin{deluxetable}{cccccccccc}
\tabletypesize{\scriptsize}
\tablecaption{Kinematic Distances to Molecular Clouds in the GRS using HISA and 21 cm continuum absorption\label{grs_table}}
\tablewidth{0pt}
\tablecolumns{10}
\tablehead{
\colhead{GRS cloud} &
 \colhead{l} &
\colhead{b} &
\colhead{$V_{LSR}$} &
\colhead{$\Delta \ell$} &
\colhead{$\Delta b$} &
\colhead{$\Delta V$}&
 \colhead{D} &
 \colhead{L} &
\colhead{21 cm cont.}\\
\colhead{   } &
\colhead{\tiny{(\degr)}} &
\colhead{\tiny{(\degr)}} &
\colhead{\tiny{(\kms)}}&
\colhead{\tiny{(\degr)}} &
\colhead{\tiny{(\degr)}} &
\colhead{\tiny{(\kms)}}&
\colhead{\tiny{(kpc)}}&
\colhead{\tiny{(10$^4$ K \kms pc$^2$)}}&
\colhead{source ?}\\
 }

\startdata
GRSMC G053.59$+$00.04 & 53.59 &   0.04 &  23.74 &   0.42 &   0.19 &   1.99 &   8.15 &   1.93 &   \\
GRSMC G029.89$-$00.06 & 29.89 &  -0.06 & 100.68 &   0.33 &   0.32 &   5.09 &   6.78 &   4.38 &  y \\
GRSMC G049.49$-$00.41 & 49.49 &  -0.41 &  56.90 &   0.26 &   0.18 &   9.77 &   5.53 &   1.54 &  y \\
GRSMC G018.89$-$00.51 & 18.89 &  -0.51 &  65.82 &   0.27 &   0.28 &   2.80 &   4.78 &   1.06 &  y \\
GRSMC G030.49$-$00.36 & 30.49 &  -0.36 &  12.26 &   0.32 &   0.26 &   4.56 &   1.00 &   0.02 &   \\
GRSMC G035.14$-$00.76 & 35.14 &  -0.76 &  35.22 &   0.30 &   0.18 &   5.00 &   2.35 &   0.33 &   \\
GRSMC G034.24$+$00.14 & 34.24 &   0.14 &  57.75 &   0.45 &   0.36 &   5.98 &   3.78 &   1.37 &  y \\
GRSMC G019.94$-$00.81 & 19.94 &  -0.81 &  42.87 &   0.32 &   0.18 &   2.81 &   3.50 &   0.38 &   \\
GRSMC G023.44$-$00.21 & 23.44 &  -0.21 & 101.10 &   0.45 &   0.41 &   5.75 &   6.43 &   4.68 &  y \\
GRSMC G038.94$-$00.46 & 38.94 &  -0.46 &  41.59 &   0.29 &   0.28 &   2.97 &  10.50 &   3.45 &   \\
GRSMC G023.44$-$00.21 & 23.44 &  -0.21 & 103.65 &   0.21 &   0.17 &   3.44 &   6.65 &   0.67 &  y \\
GRSMC G030.79$-$00.06 & 30.79 &  -0.06 &  94.73 &   0.25 &   0.31 &   6.12 &   6.22 &   3.18 &  y \\
GRSMC G030.29$-$00.21 & 30.29 &  -0.21 & 104.50 &   0.27 &   0.39 &   3.01 &   7.32 &   1.62 &   \\
GRSMC G053.14$+$00.04 & 53.14 &   0.04 &  22.04 &   0.28 &   0.33 &   2.39 &   1.80 &   0.09 &   \\
GRSMC G022.44$+$00.34 & 22.44 &   0.34 &  84.52 &   0.29 &   0.25 &   2.81 &   5.43 &   0.38 &   \\
GRSMC G024.49$+$00.49 & 24.49 &   0.49 & 102.38 &   0.46 &   0.32 &   5.24 &   6.47 &   2.82 &  y \\
GRSMC G049.39$-$00.26 & 49.39 &  -0.26 &  50.94 &   0.20 &   0.26 &   3.54 &   6.82 &   1.29 &  y \\
GRSMC G019.39$-$00.01 & 19.39 &  -0.01 &  26.72 &   0.41 &   0.23 &   3.88 &   2.42 &   0.36 &   \\
GRSMC G034.74$-$00.66 & 34.74 &  -0.66 &  46.69 &   0.28 &   0.39 &   4.33 &   3.08 &   0.66 &   \\
GRSMC G023.04$-$00.41 & 23.04 &  -0.41 &  74.32 &   0.49 &   0.29 &   4.20 &   4.90 &   1.70 &   \\
GRSMC G018.69$-$00.06 & 18.69 &  -0.06 &  45.42 &   0.26 &   0.19 &   3.86 &  12.35 &   4.00 &   \\
GRSMC G018.19$-$00.31 & 18.19 &  -0.31 &  50.09 &   0.22 &   0.28 &   4.15 &   4.07 &   0.77 &   \\
GRSMC G025.64$-$00.11 & 25.64 &  -0.11 &  93.88 &   0.32 &   0.27 &   2.78 &   5.85 &   1.11 &   \\
GRSMC G024.79$+$00.09 & 24.79 &   0.09 & 110.45 &   0.42 &   0.43 &   3.25 &   7.70 &   3.25 &  y \\
GRSMC G030.44$-$00.26 & 30.44 &  -0.26 & 103.65 &   0.36 &   0.37 &   3.83 &   7.30 &   2.37 &   \\
GRSMC G023.24$-$00.36 & 23.24 &  -0.36 &  77.30 &   0.44 &   0.33 &   2.73 &   5.05 &   1.42 &   \\
GRSMC G019.89$-$00.56 & 19.89 &  -0.56 &  44.14 &   0.32 &   0.24 &   3.90 &   3.58 &   0.61 &   \\
GRSMC G022.04$+$00.19 & 22.04 &   0.19 &  50.94 &   0.25 &   0.61 &   5.95 &   3.80 &   0.75 &   \\
GRSMC G024.39$+$00.14 & 24.39 &   0.14 & 113.43 &   0.37 &   0.34 &   4.23 &   7.72 &   4.18 &   \\
GRSMC G030.54$-$00.06 & 30.54 &  -0.06 &  93.45 &   0.31 &   0.28 &   5.13 &   6.07 &   3.36 &  y \\
\enddata

\end{deluxetable}

\clearpage

\begin{deluxetable}{cccc}
\tabletypesize{\scriptsize}
\tablecaption{Summary of the determination of the kinematic distances}
\tablewidth{0pt}
\tablecolumns{4}
\tablehead{
\colhead{} &
 \colhead{HISA} &
\colhead{HISA + 21 cm cont. source} &
\colhead{Total} \\
 }

\startdata

NEAR & 491 & 33 & 524\\
FAR & 211 & 15 & 226 \\
 & & & \\
\hline
\hline
 & & & \\
No distance & HISA inconclusive & Near-far blending & HISA and 21 cm cont. disagree \\
 & 64 & 14 & 1\\

\enddata
\label{dist_stat}
\end{deluxetable}

\begin{deluxetable}{cccccccc}
\tabletypesize{\scriptsize}
\tablecaption{Comparison between distances from maser trigonometric parallax and kinematic distances}
\tablewidth{0pt}
\tablecolumns{8}
\tablehead{
\colhead{GRS cloud} &
 \colhead{$\ell$} &
\colhead{$b$} &
\colhead{V$_{LSR}$ ($^{13}$CO)} &
 \colhead{V$_{LSR}$ (maser)} &
\colhead{D$_{kin}$} &
\colhead{D$_{parallax}$} &
\colhead{Reference}
\\
\colhead{   } &
\colhead{\tiny{(\degr)}} &
\colhead{\tiny{(\degr)}} &
\colhead{\tiny{(\kms)}}&
\colhead{\tiny{(\kms)}}&
\colhead{\tiny{(kpc)}}&
\colhead{\tiny{(kpc)}}&
 \colhead{}    \\
 }

\startdata

GRSMC G023.04-00.41 & 23.04 & -0.41 & 74.32 & 81.5 & 4.9 & 4.6 & Brunthaler et al. (2008)\\
GRSMC G023.44-00.18 & 23.44 & -0.18 & 101.1 & 97.6 & 6.5 & 5.9 & Brunthaler et al. (2008)\\
GRSMC G035.14-00.76 & 35.20 & -0.74 & 35.2 & 31.0 & 2.4 & 2.2 & Zhang et al. (2008)\\
GRSMC G049.49-00.40 & 49.49 & -0.37 & 56.9 & 56.4 & 5.5 & 5.1 & Xu et al. (2008)\\

\enddata
\label{table_reid}
\end{deluxetable}

\clearpage

\begin{figure}
\subfigure{\includegraphics[height= 7cm]{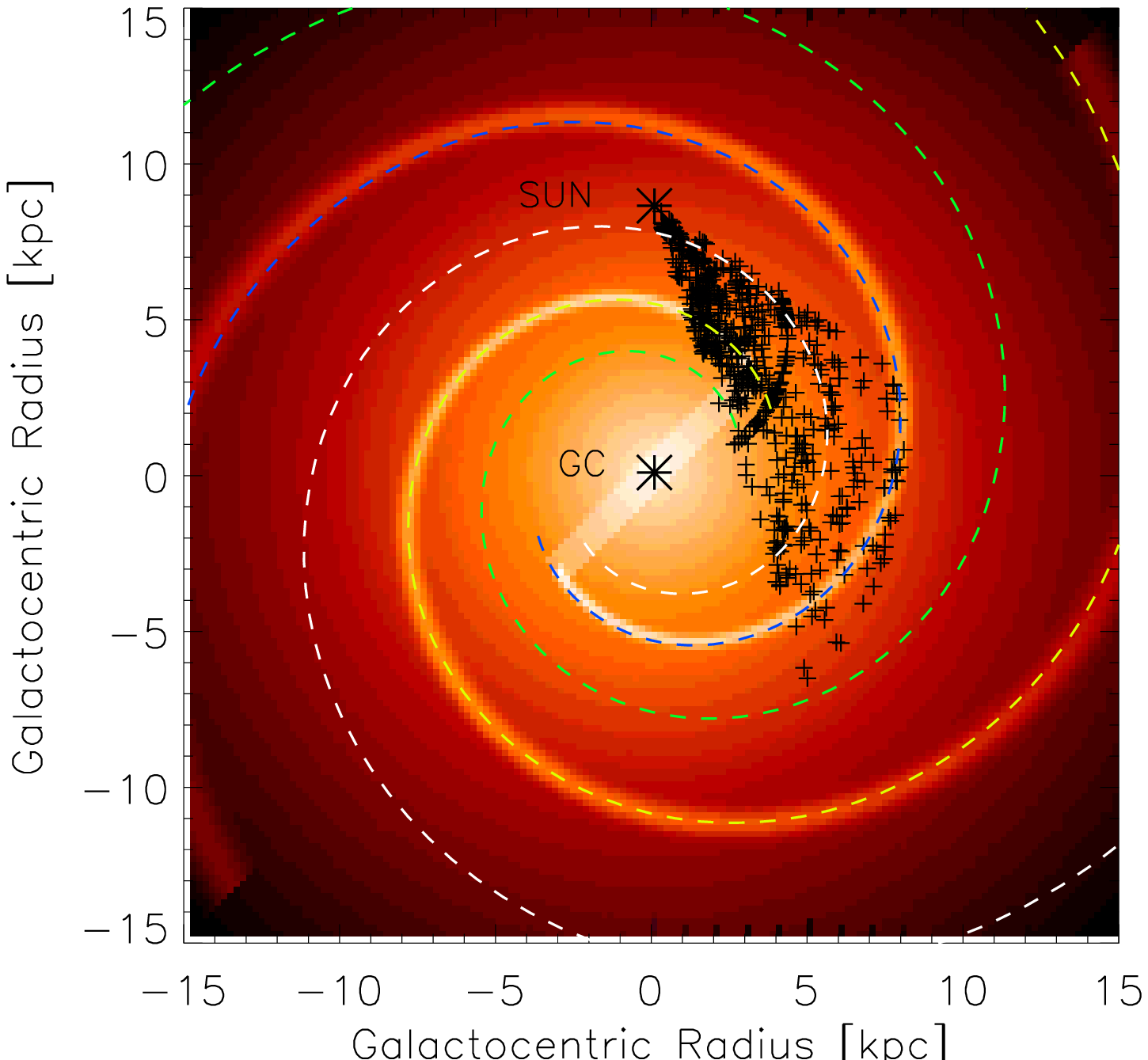}}
\subfigure{\includegraphics[height= 7cm]{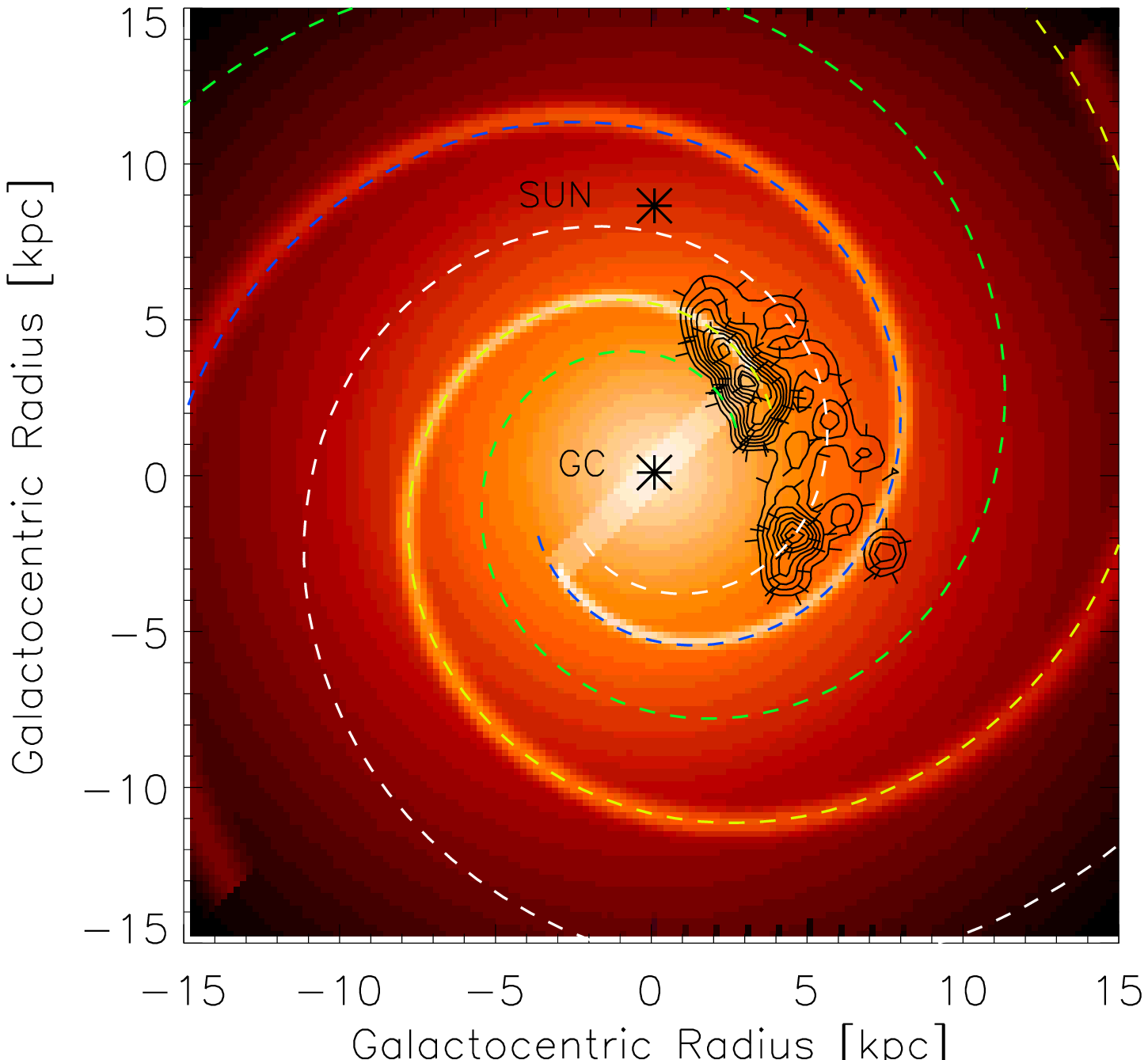}}
\subfigure{\includegraphics[height= 7cm]{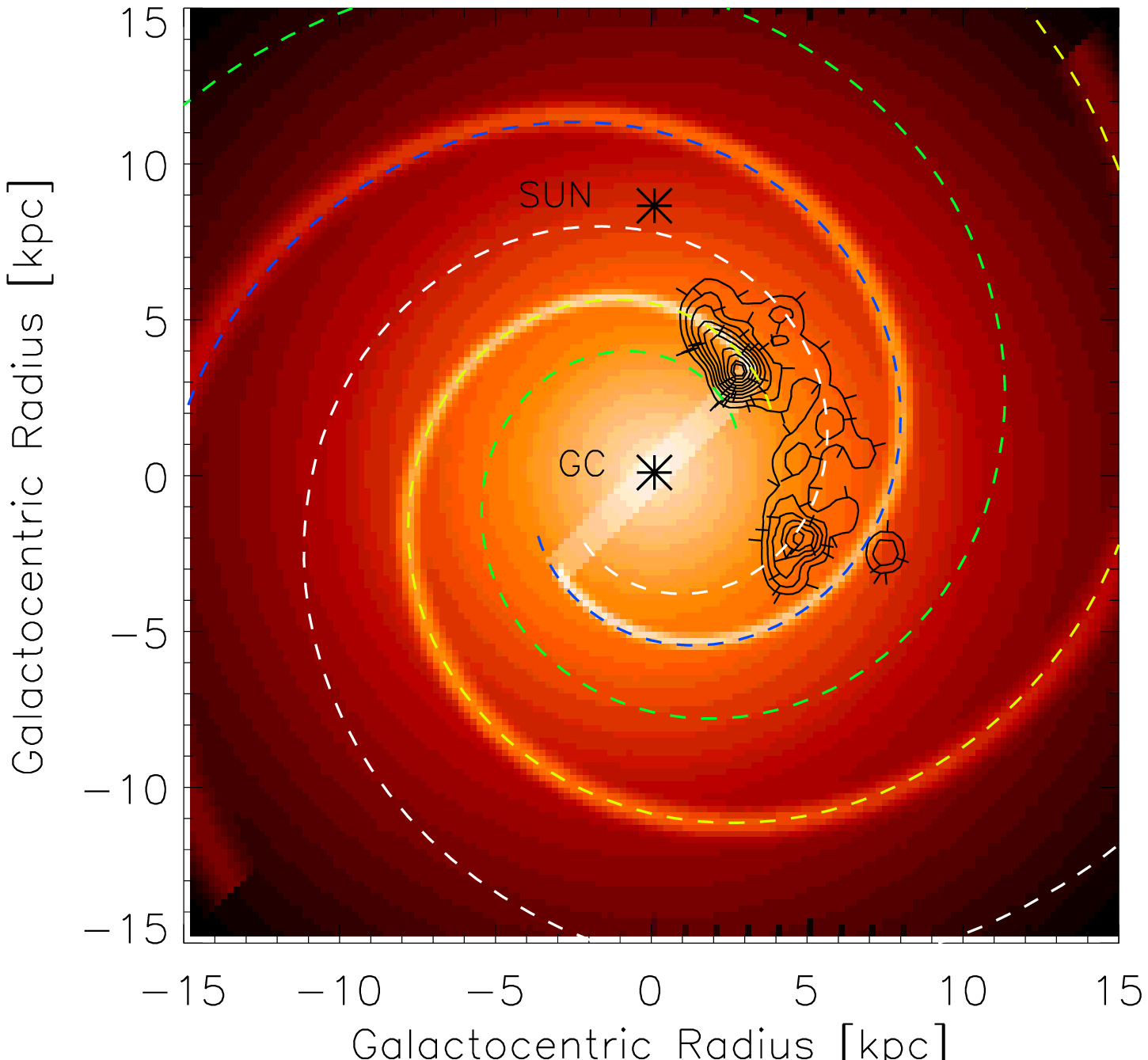}}
\caption{Top left: Location of the GRS molecular clouds (indicated by small black crosses) using kinematic distances derived from the Clemens rotation curve. The background image represents a two-arm model from the Galaxy by \citet{benjamin08} based on GLIMPSE star count data. The dashed colored lines represent the \citet{vallee95} four-arm model with a pitch angle of 12.7 \degree. The green, yellow, white and blue lines correspond to the 3 kpc-arm, the Scutum-Crux arm, the Sagittarius arm, and the Perseus arm respectively. Top right and botton: \CO surface brightness contours (in units of K \kmsn) from the GRS clouds obtained from the \citet{C85} rotation curve (top right) and a flat rotation curve scaled at V$_0$ = 220 \kms (bottom). The contour levels are 10$\%$, 20$\%$, 30$\%$, 40$\%$, 50$\%$, 60$\%$, 70$\%$, 80$\%$, 90$\%$, and 95$\%$ of the maximum surface brightness (10.1 K \kmsn). The \CO surface brightness was derived by summing the clouds' \CO luminosities over 0.04 kpc$^2$ bins, and smoothing the map by a square kernel of width 0.4 kpc.}
\label{map}
\end{figure}

\end{document}